\documentclass[
reprint,
superscriptaddress,
nobibnotes,
 amsmath,amssymb,
 aps,  physrev,
 prl,
 floatfix,
]{revtex4-2}

\usepackage{graphicx}
\usepackage{epstopdf} 
\usepackage{dcolumn}
\usepackage{bm}
\usepackage{siunitx}
\usepackage{hyperref}
\usepackage{xcolor}
\usepackage{float}

\begin{document}

\title{Ratchet motion of magnetic skyrmions driven by surface acoustic sawtooth waves}
\author{Philipp Schwenke}
\email[E-Mail: ]{philipp.schwenke@rptu.de}
\affiliation{Fachbereich Physik and Landesforschungszentrum OPTIMAS, Rheinland-Pf{\"a}lzische Technische Universit{\"a}t Kaiserslautern-Landau, 67663 Kaiserslautern, Germany}

\author{Ephraim Spindler}
\affiliation{Fachbereich Physik and Landesforschungszentrum OPTIMAS, Rheinland-Pf{\"a}lzische Technische Universit{\"a}t Kaiserslautern-Landau, 67663 Kaiserslautern, Germany}

\author{Vitaliy I. Vasyuchka}
\affiliation{Fachbereich Physik and Landesforschungszentrum OPTIMAS, Rheinland-Pf{\"a}lzische Technische Universit{\"a}t Kaiserslautern-Landau, 67663 Kaiserslautern, Germany}

\author{Alexandre Abbass Hamadeh}
\affiliation{Fachbereich Physik and Landesforschungszentrum OPTIMAS, Rheinland-Pf{\"a}lzische Technische Universit{\"a}t Kaiserslautern-Landau, 67663 Kaiserslautern, Germany}

\author{Philipp Pirro}
\affiliation{Fachbereich Physik and Landesforschungszentrum OPTIMAS, Rheinland-Pf{\"a}lzische Technische Universit{\"a}t Kaiserslautern-Landau, 67663 Kaiserslautern, Germany}

\author{Mathias Weiler}
\affiliation{Fachbereich Physik and Landesforschungszentrum OPTIMAS, Rheinland-Pf{\"a}lzische Technische Universit{\"a}t Kaiserslautern-Landau, 67663 Kaiserslautern, Germany}

\begin{abstract}
The manipulation of skyrmions by surface acoustic waves (SAW) has garnered significant interest in the field of spintronic devices. Previous studies established that skyrmions can be generated and moved by strain pulses. In this study, we propose that sawtooth-SAWs can be used to drive a ratchet motion of magnetic skyrmions in the presence of pinning centers. This results in a net motion of the skyrmions orthogonal to the continuously applied SAW. The ratchet motion is fundamentally caused by non-vanishing pinning, so that a certain strain gradient magnitude is required to overcome pinning and start skyrmion motion. We demonstrate the feasibility of our concept by micromagnetic simulations and analytical model calculations.

\end{abstract}

\maketitle

\section{\label{Introduction}Introduction\protect}
In the last decade, magnetic skyrmions have emerged as a subject of interest in the field of modern computing, particularly in the context of their potential application for data storage \cite{suess_repulsive_2018, muller_magnetic_2017, luo_skyrmion_2021, gobel_skyrmion_2021, jiang_skyrmion-based_2023}, reservoir and neuromorphic computing \cite{song_skyrmion-based_2020, pinna_reservoir_2020, raab_brownian_2022, lee_reservoir_2022, yokouchi_pattern_2022, li_magnetic_2017}.
To leverage skyrmions for these applications, it is imperative to generate and annihilate skyrmions and also control their position.
Such skyrmion transport can be achieved by employing spin-transfer or spin-orbit torques caused by electric currents~\cite{cai_current-driven_2021, zhang_magnetic_2023, purnama_guided_2015}, which is the most widespread concept, but comes with drawbacks such as Joule heating and material constraints due to requiring conductive materials, and / or large spin-orbit interactions. 
An alternative approach to manipulating skyrmions without Joule heating involves strain in the magnetic film. The application of strain by, e.g., surface acoustic waves (SAWs), enables the creation of skyrmions~\cite{chen_ordered_2023, yokouchi_creation_2020} and their manipulation through strain gradients~\cite{miyazaki_trapping_2023, liu_manipulating_2019, yang_acoustic-driven_2024, shuai_transport_2024}. 
Similar to magnetic skyrmions, the polarization of the magnetic vortex is topologically stabilized and has also already been proven to be driven by SAWs \cite{koujok_resonant_2023, iurchuk_excitation_2024, moukhader_injection_2024, seeger_experimental_2024}.
The use of SAWs is thereby desirable due to the maturity of the SAW technology and planar transducer design.
Throughout numerous studies of electrically or elastically driven skyrmion motion, the fundamental importance of pinning has been widely recognized. Additionally, research has been conducted on the impact of pinning on different types of skyrmions and by different size of a disk-shaped defect \cite{gong_skyrmion_2022} as well as on the pinning energy landscape in a real sample \cite{gruber_skyrmion_2022}. 
While pinning might seem detrimental to skyrmion motion as it increases the threshold current / strain gradient for onset of skyrmion motion, researchers have also realized that pinning can be exploited to realize deterministic skyrmion motion. For instance, the utilization of pinning centers for the creation of a guided track and the subsequent manipulation of these tracks through electrical currents have been demonstrated \cite{stosic_pinning_2017}.
Here we show that skyrmion pinning can be exploited to cause unidirectional motion of skyrmions with appropriately engineered and continuously applied SAWs. While sinusoidal SAWs would result in oscillatory motion of either pinned or unpinned skyrmions, we show that a sawtooth SAW profile can be used to generate a ratchet-like skyrmion motion, resulting in net skyrmion transport from pinning center to pinning center in a direction roughly perpendicular to the SAW propagation direction.
\begin{figure*}[ht]
\centering
\includegraphics{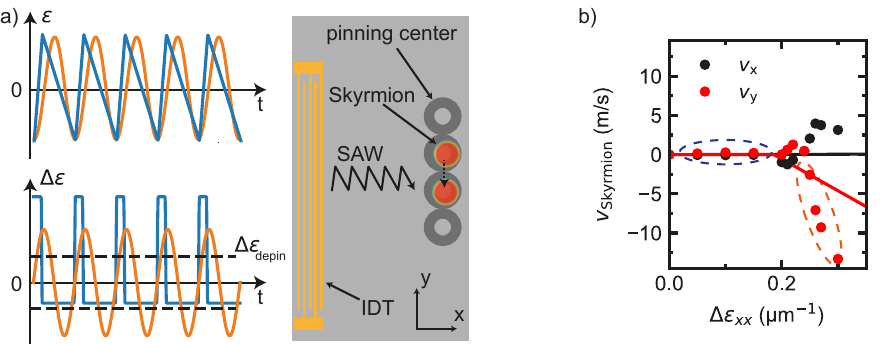}
\caption{a) Schematic of the proposed experiment. An interdigital transducer (IDT) launches a sawtooth-shaped surface acoustic wave (SAW) to a skyrmion in a pinning center. This depins the skyrmion and moves it to the next pinning center. The pinning center is modeled as a donut-shaped region with reduced anisotropy. The image on the top left shows the applied strain $\varepsilon$ and the bottom left image the resulting strain gradient $\Delta\varepsilon$ resulting from a sinusoidal-SAW (orange) and a sawtooth-shaped SAW (blue) over time $t$.
The dashed black line in the bottom left image indicates the strain gradient $\Delta\varepsilon_\text{depin}$ needed to depin the skyrmion from a pinning site.
b) depicts the micromagnetically simulated skyrmion velocities (black and red dots) of a pinned nm-scale skyrmion in a single pinning site in dependence of an applied strain gradient in x-direction. The black and red lines indicate the analytically calculated skyrmion velocity for $r_\text{SkX}=\SI{12.6}{\nano\meter}$ and $w = \SI{4.7}{\nano\meter}$ obtained from Eq.~(\ref{eq:vx}, \ref{eq:vy}).}
\label{fig:velocity}
\end{figure*}
\section{\label{Concept}Concept}
Fig.\ref{fig:velocity} a) depicts the proposed device on the right. The device consists of an interdigital transducer (IDT) that launches a sawtooth-shaped SAW towards an (arbitrary) pinning landscape, where a single skyrmion is pinned to an (initially random) pinning center. The creation of such surface acoustic waves can be experimentally implemented by the Fourier synthesis of multiple sinusoidal signals as demonstrated in~\cite{schulein_fourier_2015, weis_multiharmonic_2018}. The skyrmion can be generated by nucleation at a constriction \cite{wang_electrical_2022}, ion radiation \cite{de_jong_controlling_2023, fallon_controlled_2020, zhao_local_2023} by a strain pulse \cite{chen_ordered_2023, yokouchi_creation_2020} or electrical currents \cite{yu_current-induced_2017, akhtar_current-induced_2019, mallick_current-induced_2022, quessab_zero-field_2022, hrabec_current-induced_2017, finizio_deterministic_2019}. 
As detailed in the following, the skyrmion can be depinned from the pinning center by a sufficiently large strain gradient, while it will remain pinned for small strain gradients. We now propose to design the sawtooth profile in such a way that its rising slope results in a strain gradient that is sufficiently large to depin the skyrmion, while the falling slope is sufficiently small such that the force exerted on the skyrmion by the strain gradient stays below the depinning threshold.
In Fig. \ref{fig:velocity}a) (top‐left), the temporal evolution of the strain $\varepsilon$ is shown for both sinusoidal and sawtooth‐shaped surface acoustic waves (SAWs). On the bottom-left, the corresponding strain gradients $\Delta\varepsilon$ are plotted together with the critical gradient required for skyrmion depinning $\Delta\varepsilon_\text{depin}$. For the sinusoidal SAW, $\Delta\varepsilon$ exhibits perfect symmetry about zero, so that $|\Delta\varepsilon|$ exceeds $|\Delta\varepsilon_\text{depin}|$ during both the positive and negative half‐cycles. This symmetry confines the skyrmion to bidirectional oscillation between two neighboring pinning sites. By contrast, the sawtooth SAW produces an asymmetric $\Delta\varepsilon$ profile with a larger maximum gradient at the same applied strain amplitude, and $\Delta\varepsilon>\Delta\varepsilon_\text{depin}$ occurs only during one polarity of the waveform.
A periodic and continuously applied sawtooth SAW then depins the skyrmion and moves it from pinning center to pinning center in a unidirectional motion.
In the following we discuss the feasibility of this concept by means of analytical calculations and micromagnetic simulations.

\section{\label{Model}Analytical Model}
We start by performing an analytical calculation of the strain gradient necessary to depin a µm-sized skyrmion from a pinning center.
The magnetization of the Néel-type skyrmion in the analytical model is described by \cite{rohart_skyrmion_2013, wang_theory_2018}
\begin{align}\label{eq:m}
\begin{split}
    m_\text{x} &= M_\text{s}\cdot Q\cdot\sin(\Theta)\cos(\chi)\\
    m_\text{y} &= M_\text{s}\cdot Q\cdot\sin(\Theta)\sin(\chi)\\
    m_\text{z} &= M_\text{s}\cdot\cos(\Theta)
\end{split}
\end{align}
with $\Theta = 2\cdot\arctan\left(\exp\left(\frac{r_\text{SkX}-r}{w}\right)\right)$. $r_\text{SkX}$ describes the skyrmion radius and $r=\sqrt{x^2+y^2}$ and $\chi=\arctan(y/x)$ are the polar coordinates in relation to the Cartesian coordinate system of Fig.~\ref{fig:velocity}~a). $Q$ is the winding number and $w$ the skyrmion domain wall thickness.
With these magnetization components, the magnetoelastic free energy density contribution for a polycrystalline magnet (magnetoelastic constant $b_1=b_2$) can be written as \cite{dreher_surface_2012}
\begin{equation}
    G_\mathrm{D} = \frac{b_\text{1}}{M_\text{s}}\left(\epsilon_{xx}m_\text{x}^2+\epsilon_{zz}m_\text{z}^2+2\epsilon_{xz}m_\text{x}m_\text{z}+2\epsilon_{xy}m_\text{x}m_\text{y}\right).
    \label{eq:free-energy}
\end{equation}
In our analytical model, we assume that the skyrmion shape is rigid, viz. it does not change under the application of strain gradients. This assumption is supported by the subsequent micromagnetic simulations (see below). Moreover, we assume a uniform and linear strain gradient $\Delta\epsilon_{ij}$ over the skyrmion diameter, such as obtained by a perfect sawtooth-SAW with wavelength much larger than the skyrmion diameter. In this case, we obtain the spatial dependence of the strains as $\epsilon_{ij}(x)=\Delta\epsilon_{ij}(x+x_0)$, where $x_0$ denotes the center of the skyrmion.  By inserting this definition and Eq.~\eqref{eq:m} into Eq.~\eqref{eq:free-energy}, the spatially averaged magnetoelastic contribution to the free energy density of a skyrmion within a linear strain gradient is obtained as
\begin{align}
        G_\text{0,D} &= \frac{1}{A}\int_A dxdy \;G_\mathrm{D}
\end{align}
with the skyrmion disk area $A=\pi r_\text{SkX}^2$. Multiplying this by the volume of the skyrmion we obtain the free energy for a single skyrmion
\begin{equation}
    G_\text{0} = G_\text{0,D}\cdot 2r_\text{SkX}^2L\pi
\end{equation}
with the film thickness $L$.  We calculate the force $F_x$ acting on a skyrmion resulting from a strain gradient by differentiation of free energy with respect to the center position of the skyrmion $x_0$.
\begin{widetext}
\begin{equation}
\begin{split}
    F_x =& \frac{\text{d}G_\text{0}}{\text{d}x_\text{0}}\\ 
    =& -\frac{b_\text{1}LM_\text{s}\pi}{1+e^4}(Q^2w\Delta\epsilon_{xx}\left(2r_\text{SkX}+w\left(e^4\left(-4+\text{log}\left(1+e^4\right)\right)+\text{log}\left(1+e^4\right)\right)\right)\\&-\Delta\epsilon_{zz}    \left(\left(1+e^4\right)r_\text{SkX}^2+4\left(2+e^4\right)r_\text{SkX}w+2w^2\left(2+e^4\left(-2+\text{log}\left(1+e^4\right)\right)+\text{log}\left(1+e^4\right)\right)\right)\\&-\left(1+e^4\right)w^2\left(Q^2\Delta\epsilon_{xx}-2\Delta\epsilon_{zz}\right)\text{log}\left(1+e^{\frac{2r_\text{SkX}}{w}}\right))
    \label{eq:force_sky_ana}
\end{split}
\end{equation}
\end{widetext}

Each circular pinning potential can be defined as \cite{jiang_current-driven_2022}
\begin{equation}
    U_\text{pin} = C_\text{0}e^{-\left(r_\text{0}/R_\text{0}\right)^2}
    \label{eq:pinning-potential}
\end{equation}
with the radius $R_\text{0}$ of the pinning center and the pinning energy $C_\text{0} = M_\text{s}B_\text{pin}R_\text{0}^2L$. Here $B_\text{pin}$ denotes an effective pinning field.
We mainly focus here on the case that the skyrmion is pinned at its center. 
It is well known, that the pinning location of skyrmions can vary depending on size and that larger skyrmions tend to be pinned at the skyrmion boundary \cite{gruber_skyrmion_2022, gong_skyrmion_2022}. 
In the case that the skyrmion is pinned at its boundary the same potential but shifted by a radial coordinate $\xi$ could be used, so that $U_\text{pin} = C_\text{0}e^{-\left(|r_\text{0}-\xi|/R_\text{0}\right)^2}$ \cite{jiang_current-driven_2022}.
From Eq.~(\ref{eq:pinning-potential}) the maximum pinning force can be expressed as 
\begin{equation}
    F_\text{pin,max} = B_\text{pin}\sqrt{\frac{2}{e}} L M_\text{s} R_\text{0}.
    \label{eq:Fpinmax}
\end{equation}
The skyrmion velocity in dependence on the applied strain can be calculated by \cite{liu_manipulating_2019}
\begin{align}
    v_{x} = \frac{\alpha D}{\alpha^2D^2+G^2}\cdot\left(F_{x}-\text{sign}(F_{x})F_\text{pin}\right) \label{eq:vx}\\
    v_{y} = -\frac{G}{\alpha^2D^2+G^2}\cdot\left(F_{x}-\text{sign}(F_{x})F_\text{pin}\right)
    \label{eq:vy}
\end{align}
with $G = \frac{4\pi QM_\text{s}L}{\gamma}$, $D = |{G}|$ and $\alpha$ the gilbert damping.
The resulting calculated skyrmion velocity at the location of maximum pinning in dependence of an applied strain gradient for $r_\text{SkX} = \SI{12.6}{\nano\meter}$, $w = \SI{4.7}{\nano\meter}$, $b_1 = \SI{-8}{\tesla}$, $\lambda_\text{SAW} = \SI{1}{\micro\meter}$, $R_\text{0} = \SI{25.5}{\nano\meter}$, $\alpha=0.01$ and $B_\text{pin} = \SI{24.6}{\milli\tesla}$ is shown in Fig.~\ref{fig:velocity} b) as black ($v_\text{x}$) and red ($v_\text{y}$) lines. 
The radius and domain wall width is determined by fitting $m_\text{z}$ from Eq.~(\ref{eq:m}) to the magnetization profile from the skyrmion in the micromagnetic simulation.
The value of $B_\text{pin}$ and $R_\text{0}$ is determined by fitting the pinning potential from Eq.~(\ref{eq:pinning-potential}) to the energy landscape of the pinning region in the subsequent micromagnetic simulations (see below).
At low strain gradients (blue dashed ellipse) the skyrmions velocity stays at 0, meaning that the skyrmion is pinned on the pinning center and can not move away form it.
At higher applied strain gradients (orange dashed ellipse) the velocity in y-direction increases while the velocity in x-direction stays at 0.
This can be explained by the fact that the force the skyrmion experiences in x-direction is not sufficient to depin the skyrmion in this direction.
In reality and micromagnetic simulations the skyrmion would move away from the pinning center and thus also have a non zero component of the velocity in x-direction.
We assume a purely longitudinal strain wave ($\epsilon_{xx} \neq 0$) as a simple approximation \cite{morgan_surface_2010}. To overcome the pinning, $F_{x}$ needs to be larger than $F_\text{pin,max}$ and the minimal strain needed to depin the skyrmion is given by
\begin{widetext}
\begin{equation}
    \Delta\epsilon_{xx\text{, min}} = \frac{B_\text{pin}\sqrt{\frac{2}{e}}\left(1+e^4\right)R_\text{0}}{b_\text{1}\pi Q^2w\left(2r_\text{SkX}+w\left(e^4\left(-4+\text{log}\left(1+e^4\right)\right)+\text{log}\left(1+e^4\right)\right)-\left(1+e^4\right)w\cdot\text{log}\left(1+e^{\frac{2r_\text{SkX}}{w}}\right)\right)}
    \label{eq:minstrain}
\end{equation}
\end{widetext}
With the parameters stated above, the strain needed to depin a skyrmion is $\epsilon_{{xx}_\text{, min}} = \SI{0.18}{}$.
Dividing this by $\lambda_\text{SAW}$ we get the required strain gradient $\Delta\epsilon = \SI{0.18}{\per\micro\meter}$.
A comparison of the strain required to depin a skyrmion with experimentally accessible strains generated by surface acoustic waves (SAWs) reveals that the application of the latter is indeed feasible \cite{weiler_elastically_2011, weiler_spin_2012}.
Furthermore, the pinning energy and, therefore, the required strain could be further reduced by periodic field excitations \cite{gruber_300-times-increased_2023}.

\section{\label{Simulation}Micromagnetic Simulation}
To verify that the analytical calculations resemble the expected behavior of a skyrmion, and to demonstrate that the concept can be transferred also to nanoscale skyrmions, micromagnetic simulations on nm-sized skyrmions are performed using mumax3 \cite{vansteenkiste_design_2014} within the Aithericon framework \cite{noauthor_see_nodate}.
We use a cell size of $1\times 1\times 1$ \si{\nano\metre\cubed} and a grid size of $256\times 256$ with periodic boundary conditions.
The temperature is set to $T = 0$.
The CoFeB sample parameters for the simulation were taken from R. Gruber \textit{et.~al} \cite{gruber_skyrmion_2022}.
The saturation magnetization is set to $M_\text{s} =  \SI{1}{\mega\ampere\per\metre}$, the exchange stiffness as $A_\text{ex} =~\SI{15}{\pico\joule\per\metre}$, and the Gilbert damping as $\alpha = 0.01$. 
The anisotropy is set to $K_\text{u} = \SI{1.1}{\mega\joule\per\meter\cubed}$.
The magnetoelastic coupling is $B_\text{1}= B_\text{2} =b_\text{1}\cdot M_\text{s} =\SI{-8}{\mega\joule\per\meter\cubed}$ and the interfacial Dzyaloshinskii-Moriya interaction strength $D = \SI{3.217}{\milli\joule\per\meter\squared}$.
\begin{figure}[ht]
\centering
\includegraphics{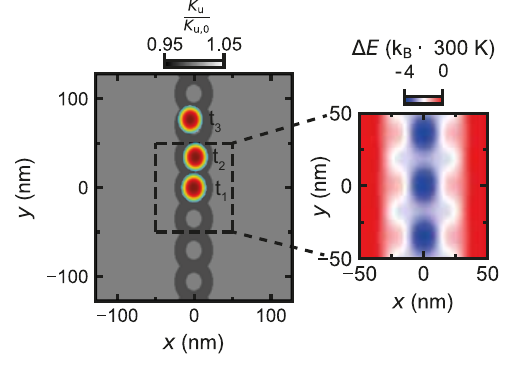}
\caption{Anisotropy landscape used for simulating the response of the skyrmion to a surface-acoustic-wave. The anisotropy is reduced by a value of $2$~\% in a donut shaped form. The regions of reduced anisotropy create a chain like structure. The anisotropy landscape creates a energy landscape which induces pinning to the system. The red circles correspond to skyrmions in the landscape after $t_\text{1} = \SI{0}{\nano\second}$, $t_\text{2} = \SI{30}{\nano\second}$ and $t_\text{3} = \SI{40}{\nano\second}$ after applying a sawtooth-shaped SAW. The energy landscape for a region of $100\times100$ \si{\nano\metre} is shown in on the right.}
\label{fig:chain-pinning}
\end{figure}
A Neel-type skyrmion is initialized in the center of the simulation region.
A magnetic field of $\mu_0H = \SI{0.05}{\tesla}$ in the out-of plane direction is applied to stabilize the skyrmion.
The resulting skyrmion has a size of approximately $30$ nm.
At the region where the skyrmion is initialized, the anisotropy is reduced by 2\% in a donut-shaped form to create a pinning center by reducing the energy of the skyrmion domain wall in this region.
It should be noted that the shape of this pinning center is chosen arbitrarily as the minimum strain gradient needed to depin is solely dependent on the skyrmion size and maximum pinning force.
The strain gradients are then systematically varied to determine the strain required to depin the skyrmion.
In Fig.~\ref{fig:velocity} c), the velocity of the skyrmion in the x and y directions is plotted against a constant strain gradient in the x direction, with the skyrmion located within the center of the pinning site together with the analytically calculated velocity indicated by the red ($v_\text{y}$) and black ($v_\text{x}$) lines.
In qualitative agreement with the analytically calculated velocities (compare the red and blue lines in Fig.~\ref{fig:velocity} b), the velocity stays approximately 0 for low applied strain gradients (blue dashed ellipse) and the skyrmion is depinned and can move freely at a sufficiently large applied strain gradient (orange dashed ellipse).
\begin{figure*}[ht]
\centering
\includegraphics{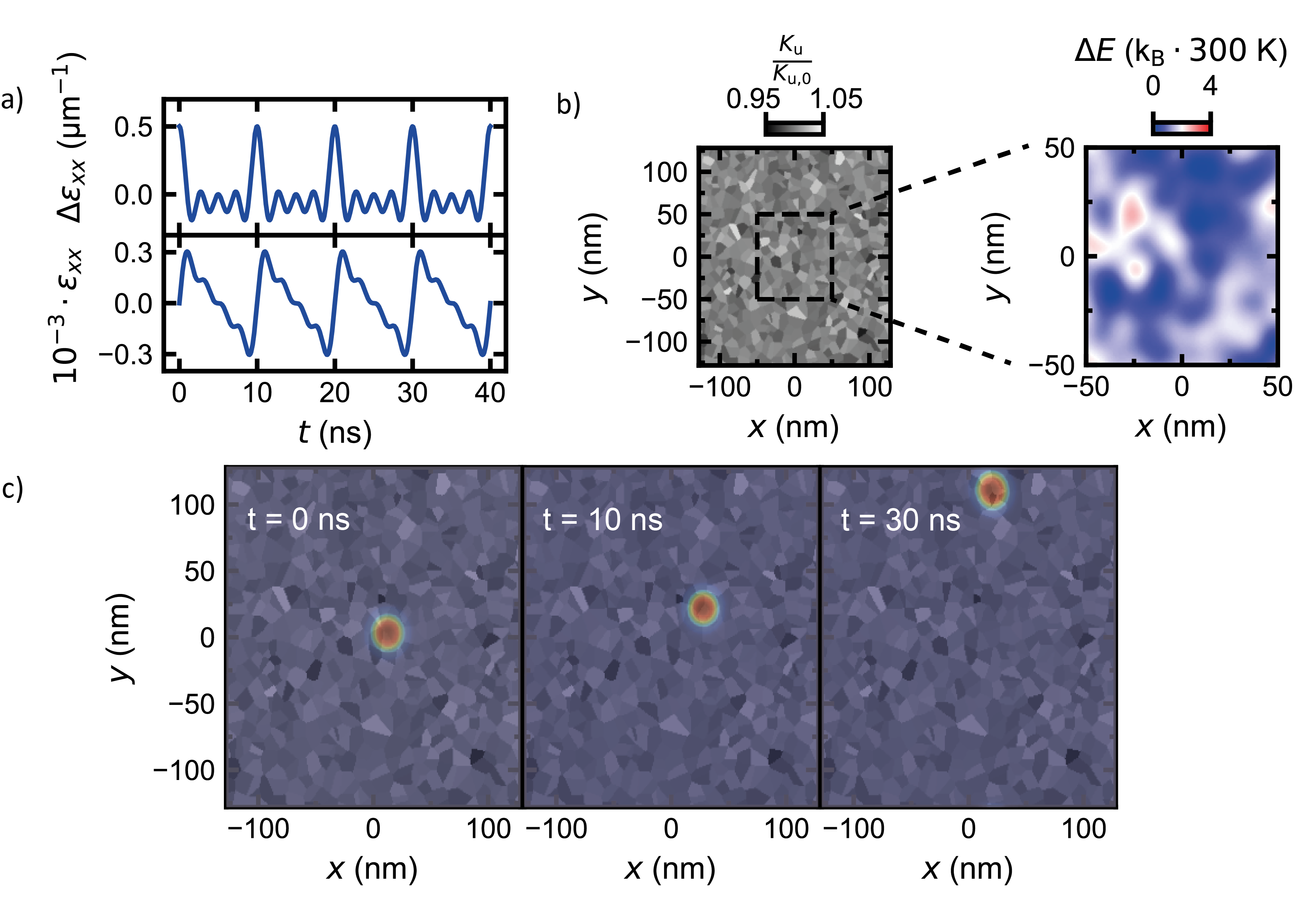}
\caption{a) Shape of the resulting strain $\epsilon_{xx}$ (bottom) and the straingradient $\Delta\epsilon_{xx}$ (top) due to the applied sawtooth SAW. b) Anisotropy landscape created by $\SI{10}{\nano\meter}$ sized grains with random anisotropy reduction between 0 and $1\%$. On the right the resulting energy landscape in a $100\times100$ $\si{\nano\meter\squared}$ region. c) Movement of the skyrmion placed in the anisotropy landscape with an sawtooth-shaped SAW applied in the positive x-direction after a time of $ t = \SI{0}{\nano\second}$, $ t = \SI{10}{\nano\second}$ and $ t = \SI{30}{\nano\second}$.}
\label{fig:Sawtoothperiod-ani}
\end{figure*}
The discrepancy in the skyrmion velocity of the depinned skyrmion in the micromagentic simulations compared to the analytically calculated velocity can be explained by the fact that in the micromagnetic simulations the skyrmion can move away from the pinning site, while in the analytical model the velocity is always calculated at the location of maximum pinning.
This results in the skyrmion velocity being significantly smaller than in the unpinned case.
Thus, it is evident from this figure that the strain gradient must exceed a threshold to facilitate skyrmion depinning from the region of reduced anisotropy.
Subsequent to depinning, the skyrmion velocity increases to a value equivalent to that observed in the absence of a pinning center.
It additionally shows that the skyrmion does not move fully perpendicular to the applied strain but at an angle due to the skyrmion Hall effect \cite{kim_asymmetric_2018, woo_current-driven_2018, brearton_deriving_2021, jiang_direct_2017, yang_fundamentals_2024, chen_skyrmion_2017, litzius_skyrmion_2017}. 

However, it should be noted that part of the velocity observed in the x-direction arises from the method used to evaluate the velocities.
We determine the skyrmion position at $t=\SI{0}{\nano\second}$ and again at $t=\SI{1}{\nano\second}$, after applying the constant strain gradient.
As a result, motion of the skyrmion due to the pinning center is included and contributes to the measured velocity.
In the analytical calculations, by contrast, the x-velocity is nearly zero, since the pinning center does not affect the velocity evaluation.
Because we initialize the skyrmion and allow it to relax before applying the strain gradient, and only evaluate the position after sufficient time for the skyrmion to depin under a large enough strain gradient, we did not perform statistical averaging of the velocities. Our focus is solely on determining the minimum strain gradient, $\Delta\epsilon_{xx\text{, min}}$, required to depin the skyrmion.
In the Supplementary Material, this is also demonstrated for a purely circular pinning center \cite{supplement}.

In order to ascertain the feasibility of relocating a skyrmion from one pinning site to another via an applied strain gradient, the single-pinning center is replicated to create a pinning chain.
The resulting pinning chain, in conjunction with the energy landscape, is shown in Fig.~\ref{fig:chain-pinning}. 
The anisotropy is reduced by 2\% in a donut-shaped manner to form each pinning center.
The right side of Fig.~\ref{fig:chain-pinning} presents the energy landscape resulting from the anisotropy region.
The configuration with the lowest energy is that in which the skyrmion domain wall region matches the region of reduced anisotropy.
Given that the minimum energy regions are along the chain at the pinning centers, the skyrmion is inclined to move along the chain direction when applying a strain in the x-direction as the gradient needed to move it perpendicular to the chain would be higher.
When a sufficiently high strain gradient is applied in the x-direction, the skyrmion can be depinned and move towards the next pinning center.
This is depicted in Fig.~\ref{fig:chain-pinning} by the skyrmion at times $t_\text{2} = \SI{30}{\nano\second}$ and $t_\text{3} = \SI{40}{\nano\second}$ after nucleating it in the center at $t_\text{1} = \SI{0}{\nano\second}$ and applying a sawtooth-shaped SAW with a frequency of $f = \SI{100}{\mega\hertz}$ and displacement only in the x-direction \cite{morgan_surface_2010}.
This is realized by a time varying strain gradient with the assumption that $\lambda_\text{SAW}>>r_\text{SkX}$.
The full movement of the skyrmion over time is shown in the Supplementary Material \cite{supplement}.
In the case of applying the strain gradient in the opposite direction, the skyrmion also depins and moves in the opposite direction towards the next pinning center.
Thus, if a sinusoidal surface acoustic wave is applied, the skyrmion would end up at the starting position due to the symmetric strain profile.
To show this, we demonstrate the movement of the skyrmion for an applied sinusoidal as well as a triangular-shaped SAW, which also has a symmetric strain profile, in the supplementary material \cite{supplement}.
This can be overcome by utilization of a sawtooth-shaped surface acoustic wave for the movement of the skyrmion.
The creation of such a sawtooth-shaped SAW on a sample can be achieved through Fourier synthesis and the utilization of split-52 IDTs \cite{schulein_fourier_2015, weis_multiharmonic_2018}.
The resulting strain $\epsilon_{xx}$ and strain gradient $\Delta\epsilon_{xx}$ from the sawtooth SAW, created by a Fourier synthesis of 4 sine waves, is shown in Fig.~\ref{fig:Sawtoothperiod-ani} a).
In this configuration, the maximum strain gradient in the positive and negative directions is distinct.
Consequently, the depinning of skyrmions is possible during the rising slope of the sawtooth SAW where the strain gradient is strong. 
On the falling slope of each SAW cycle, the skyrmions stay pinned.

Given that a chain of donut-shaped pinning centers does not accurately represent the actual pinning sites in a real sample, we implemented a more realistic pinning landscape. To this end, a randomized region with a grain size of 10 nm was utilized. The anisotropy within the grains was randomly reduced between 0 and 1\% of $K_\text{u}$. 
The resulting anisotropy landscape is illustrated in Fig.~\ref{fig:Sawtoothperiod-ani} b).
From the anisotropy landscape, we calculated the energy landscape resulting from the regions of reduced anisotropy.
This is illustrated in Fig.~\ref{fig:Sawtoothperiod-ani} b) on the right, which presents the energy landscape for a region of $100\times100\text{ }\si{\nano\meter\squared}$ in the center. 
It is evident from the figure that the energy landscape now exhibits no discernible trajectory that would predetermine the skyrmion path by geometry of the simulation setup.
The employed energy landscape is in good quantitative agreement with the experimentally recorded energy landscapes in R.Gruber~\textit{et.~al}~\cite{gruber_skyrmion_2022} and consequently, we believe that our artificial energy landscape is a reasonable model for pinning in real samples.
In Fig.~\ref{fig:Sawtoothperiod-ani} c), a skyrmion is positioned at the center of the anisotropy landscape depicted in Fig.\ref{fig:Sawtoothperiod-ani} b) at a time designated as $t = 0$. Then, a sawtooth-SAW is applied to the simulation area. 
This is achieved by applying a time-dependent constant strain gradient over the simulation area, in accordance to the most likely experimental implementation, where the SAW wavelength is considerably larger than the simulated area. It should be noted that only $\epsilon_{xx} \neq 0$. After an initial 10-nanosecond period and the first cycle of the sawtooth-SAW (refer to Fig.~\ref{fig:Sawtoothperiod-ani} a), the skyrmion undergoes a slight displacement in the positive x and y directions.
After $t = \SI{30}{\nano\second}$ the skyrmion reaches the top boundary of the simulated area while the x-position stayed approximately constant.
In the aforementioned time periods, the skyrmion demonstrates notable movement when a high strain is applied, while remaining relatively stationary under low strain conditions leading to a ratchet motion of the skyrmion under an applied sawtooth SAW.
The full time dependence of the y-position of the skyrmion under the influence of a sawtooth-shaped, sinusoidal and triangular SAW is shown in the Supplementary Material \cite{supplement}, as well as the mean movement of a skyrmion placed in multiple different random pinning landscapes and applying a sawtooth-shaped or sinusoidal SAW. 

In our simulation the strain applied on the sample is $\epsilon_{xx} = 0.3\cdot10^{-3}$.
To verify that the proposed experiment is feasible we compare the strain used in the experiment with the maximum achievable strain before sample destruction.
This is for many materials generally in the order of $0.1\%$ to $1\%$ and given by the relation between the Young's modulus and yield strength \cite{ashby_chapter_2011}.
Moreover, SAW amplitudes of about $\SI{8}{\nano\meter}$ at a distance of $\SI{4.5}{\milli\meter}$ from the IDT and with a wavelength of $\SI{20}{\micro\meter}$ corresponding to a strain of $\epsilon = 2\cdot10^{-4}$ have been reported \cite{kavalerov_observation_2000}.
Taking into account that the reported value was not measured directly at the IDT we believe that it is possible to recreate the ratchet-like motion shown on our simulation.
Moreover, considering that the required strain to depin the skyrmion is inversely proportional to the skyrmion radius $r_\text{SkX}$ and domain wall width $w$ (compare Eq.~(\ref{eq:minstrain})) a low strain amplitude in the $10^{-6}$ range is according to the analytical calculation sufficient to depin \textmu m sized skyrmions.
Thus, the strain needed to depin the skyrmion significantly decreases when using larger skyrmions.

\begin{figure}[h]
\centering
\includegraphics{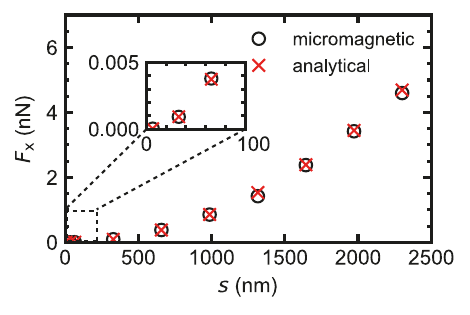}
\caption{Force on the skyrmion in dependence of the skyrmion size $s = 2\cdot r_\text{SkX}+ w$ obtained from micromagnetic simulations (black circles) and from an analytical calculation from Eq.~\eqref{eq:force_sky_ana}. $r_\text{SkX}$ and $w$ for the analytical calculation are determined by fitting the normalized $m_z$ component from Eq.~\eqref{eq:m} to the skyrmion profile of the micromagnetically simulated skyrmion.
The inset is a zoom in on small skyrmion sizes $s$.}
\label{fig:Analytic}
\end{figure}

In order to verify the compatibility between the analytical calculations and the micromagnetic simulations, we simulate the force from the strain gradient acting on the skyrmion as a function of skyrmion size for $\Delta\epsilon_{xx} = \SI{0.25}{\per\micro\meter}$, $b_\text{1} = \SI{8}{\tesla}$ and $L=\SI{1}{\nano\meter}$.
In the micromagnetic simulations the force is determined by placing the skyrmion without relaxing it at various x-positions of the simulation area and reading out the energy of the skyrmion. 
The gradient of the energy corresponds then to the force acting on the skyrmion from the strain gradient.
The micromagnetically simulated skyrmion profile is fitted to Eq.~\eqref{eq:m} in order to extract $r_\text{SkX}$ and $w$. These parameters, together with the values given above, are then inserted into Eq.~\eqref{eq:force_sky_ana} to calculate the force acting on the skyrmion.
The results are presented in Fig.~\ref{fig:Analytic}.
Here $s$ is the skyrmion size, defined as $s = 2\cdot r_\text{SkX}+ w$, since both $r_\text{SkX}$ and $w$ vary with skyrmion size in the micromagnetic simulations.
The black circles represent the force acting on the skyrmion as obtained from micromagnetics, while the red X corresponds to the analytically calculated force using Eq.~(\ref{eq:force_sky_ana})
The inset illustrates the data for small skyrmion sizes.
The results demonstrate good agreement between the analytical calculation and the micromagnetically simulated data for the force acting on the skyrmion under the applied strain gradient. 
Owing to this agreement between the micromagnetic simulations and the analytical model for nanoscale skyrmions, we conclude that our analytical model can reliably be used to estimate the strain gradient required to depin rigid skyrmions

\section{\label{Conclusion}Conclusion\protect}

In summary, we have shown by analytical model calculations as well as micromagnetic simulations, that sawtooth-shaped surface acoustic waves (SAWs) can be employed to move magnetic skyrmions in a ratchet-like motion from pinning center to pinning center. We estimate the strains required to move nanoscale or microscale skyrmions and find that the generation of the required strains should be experimentally feasible.

\begin{acknowledgments}
\section{Acknowldegements}
We acknowledge financial support by the Deutsche Forschungsgemeinschaft (DFG, German Research Foundation) - Project No. 403505631.
We acknowledge financial support from the European Research Council (ERC) under the European Union’s Horizon Europe research and innovation programme (Grant agreement No. 101044526).
A.H. and P.P. acknowledge support by the Deutsche Forschungsgemeinschaft (DFG, German Research Foundation) - TRR 173-268565370 (“Spin + X”, Project B01).

\textit{Data availability}: The data that support the findings of
 this article are openly available \cite{schwenke_ratchet_2025}, embargo periods may
 apply.

\end{acknowledgments}


\begin{thebibliography}{59}%
\makeatletter
\providecommand \@ifxundefined [1]{%
 \@ifx{#1\undefined}
}%
\providecommand \@ifnum [1]{%
 \ifnum #1\expandafter \@firstoftwo
 \else \expandafter \@secondoftwo
 \fi
}%
\providecommand \@ifx [1]{%
 \ifx #1\expandafter \@firstoftwo
 \else \expandafter \@secondoftwo
 \fi
}%
\providecommand \natexlab [1]{#1}%
\providecommand \enquote  [1]{``#1''}%
\providecommand \bibnamefont  [1]{#1}%
\providecommand \bibfnamefont [1]{#1}%
\providecommand \citenamefont [1]{#1}%
\providecommand \href@noop [0]{\@secondoftwo}%
\providecommand \href [0]{\begingroup \@sanitize@url \@href}%
\providecommand \@href[1]{\@@startlink{#1}\@@href}%
\providecommand \@@href[1]{\endgroup#1\@@endlink}%
\providecommand \@sanitize@url [0]{\catcode `\\12\catcode `\$12\catcode `\&12\catcode `\#12\catcode `\^12\catcode `\_12\catcode `\%12\relax}%
\providecommand \@@startlink[1]{}%
\providecommand \@@endlink[0]{}%
\providecommand \url  [0]{\begingroup\@sanitize@url \@url }%
\providecommand \@url [1]{\endgroup\@href {#1}{\urlprefix }}%
\providecommand \urlprefix  [0]{URL }%
\providecommand \Eprint [0]{\href }%
\providecommand \doibase [0]{https://doi.org/}%
\providecommand \selectlanguage [0]{\@gobble}%
\providecommand \bibinfo  [0]{\@secondoftwo}%
\providecommand \bibfield  [0]{\@secondoftwo}%
\providecommand \translation [1]{[#1]}%
\providecommand \BibitemOpen [0]{}%
\providecommand \bibitemStop [0]{}%
\providecommand \bibitemNoStop [0]{.\EOS\space}%
\providecommand \EOS [0]{\spacefactor3000\relax}%
\providecommand \BibitemShut  [1]{\csname bibitem#1\endcsname}%
\let\auto@bib@innerbib\@empty
\bibitem [{\citenamefont {Suess}\ \emph {et~al.}(2018)\citenamefont {Suess}, \citenamefont {Vogler}, \citenamefont {Bruckner}, \citenamefont {Heistracher},\ and\ \citenamefont {Abert}}]{suess_repulsive_2018}%
  \BibitemOpen
  \bibfield  {author} {\bibinfo {author} {\bibfnamefont {D.}~\bibnamefont {Suess}}, \bibinfo {author} {\bibfnamefont {C.}~\bibnamefont {Vogler}}, \bibinfo {author} {\bibfnamefont {F.}~\bibnamefont {Bruckner}}, \bibinfo {author} {\bibfnamefont {P.}~\bibnamefont {Heistracher}},\ and\ \bibinfo {author} {\bibfnamefont {C.}~\bibnamefont {Abert}},\ }\bibfield  {title} {\bibinfo {title} {A repulsive skyrmion chain as a guiding track for a racetrack memory},\ }\href {https://doi.org/10.1063/1.4993957} {\bibfield  {journal} {\bibinfo  {journal} {AIP Adv.}\ }\textbf {\bibinfo {volume} {8}},\ \bibinfo {pages} {115301} (\bibinfo {year} {2018})}\BibitemShut {NoStop}%
\bibitem [{\citenamefont {Müller}(2017)}]{muller_magnetic_2017}%
  \BibitemOpen
  \bibfield  {author} {\bibinfo {author} {\bibfnamefont {J.}~\bibnamefont {Müller}},\ }\bibfield  {title} {\bibinfo {title} {Magnetic skyrmions on a two-lane racetrack},\ }\href {https://doi.org/10.1088/1367-2630/aa5b55} {\bibfield  {journal} {\bibinfo  {journal} {New J. Phys.}\ }\textbf {\bibinfo {volume} {19}},\ \bibinfo {pages} {025002} (\bibinfo {year} {2017})}\BibitemShut {NoStop}%
\bibitem [{\citenamefont {Luo}\ and\ \citenamefont {You}(2021)}]{luo_skyrmion_2021}%
  \BibitemOpen
  \bibfield  {author} {\bibinfo {author} {\bibfnamefont {S.}~\bibnamefont {Luo}}\ and\ \bibinfo {author} {\bibfnamefont {L.}~\bibnamefont {You}},\ }\bibfield  {title} {\bibinfo {title} {Skyrmion devices for memory and logic applications},\ }\href {https://doi.org/10.1063/5.0042917} {\bibfield  {journal} {\bibinfo  {journal} {APL Mater.}\ }\textbf {\bibinfo {volume} {9}},\ \bibinfo {pages} {050901} (\bibinfo {year} {2021})}\BibitemShut {NoStop}%
\bibitem [{\citenamefont {Göbel}\ and\ \citenamefont {Mertig}(2021)}]{gobel_skyrmion_2021}%
  \BibitemOpen
  \bibfield  {author} {\bibinfo {author} {\bibfnamefont {B.}~\bibnamefont {Göbel}}\ and\ \bibinfo {author} {\bibfnamefont {I.}~\bibnamefont {Mertig}},\ }\bibfield  {title} {\bibinfo {title} {Skyrmion ratchet propagation: utilizing the skyrmion {Hall} effect in {AC} racetrack storage devices},\ }\href {https://doi.org/10.1038/s41598-021-81992-0} {\bibfield  {journal} {\bibinfo  {journal} {Sci. Rep.}\ }\textbf {\bibinfo {volume} {11}},\ \bibinfo {pages} {3020} (\bibinfo {year} {2021})}\BibitemShut {NoStop}%
\bibitem [{\citenamefont {Jiang}\ \emph {et~al.}(2023)\citenamefont {Jiang}, \citenamefont {Yu},\ and\ \citenamefont {Chen}}]{jiang_skyrmion-based_2023}%
  \BibitemOpen
  \bibfield  {author} {\bibinfo {author} {\bibfnamefont {Y.}~\bibnamefont {Jiang}}, \bibinfo {author} {\bibfnamefont {H.}~\bibnamefont {Yu}},\ and\ \bibinfo {author} {\bibfnamefont {X.}~\bibnamefont {Chen}},\ }\bibfield  {title} {\bibinfo {title} {Skyrmion-based racetrack multilevel data storage device manipulated by pinning},\ }\href {https://doi.org/10.1063/5.0151304} {\bibfield  {journal} {\bibinfo  {journal} {J. Appl. Phys.}\ }\textbf {\bibinfo {volume} {134}},\ \bibinfo {pages} {053901} (\bibinfo {year} {2023})}\BibitemShut {NoStop}%
\bibitem [{\citenamefont {Song}\ \emph {et~al.}(2020)\citenamefont {Song}, \citenamefont {Jeong}, \citenamefont {Pan}, \citenamefont {Zhang}, \citenamefont {Xia}, \citenamefont {Cha}, \citenamefont {Park}, \citenamefont {Kim}, \citenamefont {Finizio}, \citenamefont {Raabe}, \citenamefont {Chang}, \citenamefont {Zhou}, \citenamefont {Zhao}, \citenamefont {Kang}, \citenamefont {Ju},\ and\ \citenamefont {Woo}}]{song_skyrmion-based_2020}%
  \BibitemOpen
  \bibfield  {author} {\bibinfo {author} {\bibfnamefont {K.~M.}\ \bibnamefont {Song}}, \bibinfo {author} {\bibfnamefont {J.-S.}\ \bibnamefont {Jeong}}, \bibinfo {author} {\bibfnamefont {B.}~\bibnamefont {Pan}}, \bibinfo {author} {\bibfnamefont {X.}~\bibnamefont {Zhang}}, \bibinfo {author} {\bibfnamefont {J.}~\bibnamefont {Xia}}, \bibinfo {author} {\bibfnamefont {S.}~\bibnamefont {Cha}}, \bibinfo {author} {\bibfnamefont {T.-E.}\ \bibnamefont {Park}}, \bibinfo {author} {\bibfnamefont {K.}~\bibnamefont {Kim}}, \bibinfo {author} {\bibfnamefont {S.}~\bibnamefont {Finizio}}, \bibinfo {author} {\bibfnamefont {J.}~\bibnamefont {Raabe}}, \bibinfo {author} {\bibfnamefont {J.}~\bibnamefont {Chang}}, \bibinfo {author} {\bibfnamefont {Y.}~\bibnamefont {Zhou}}, \bibinfo {author} {\bibfnamefont {W.}~\bibnamefont {Zhao}}, \bibinfo {author} {\bibfnamefont {W.}~\bibnamefont {Kang}}, \bibinfo {author} {\bibfnamefont {H.}~\bibnamefont {Ju}},\ and\ \bibinfo {author} {\bibfnamefont {S.}~\bibnamefont {Woo}},\ }\bibfield  {title}
  {\bibinfo {title} {Skyrmion-based artificial synapses for neuromorphic computing},\ }\href {https://doi.org/10.1038/s41928-020-0385-0} {\bibfield  {journal} {\bibinfo  {journal} {Nat. Electron.}\ }\textbf {\bibinfo {volume} {3}},\ \bibinfo {pages} {148} (\bibinfo {year} {2020})}\BibitemShut {NoStop}%
\bibitem [{\citenamefont {Pinna}\ \emph {et~al.}(2020)\citenamefont {Pinna}, \citenamefont {Bourianoff},\ and\ \citenamefont {Everschor-Sitte}}]{pinna_reservoir_2020}%
  \BibitemOpen
  \bibfield  {author} {\bibinfo {author} {\bibfnamefont {D.}~\bibnamefont {Pinna}}, \bibinfo {author} {\bibfnamefont {G.}~\bibnamefont {Bourianoff}},\ and\ \bibinfo {author} {\bibfnamefont {K.}~\bibnamefont {Everschor-Sitte}},\ }\bibfield  {title} {\bibinfo {title} {Reservoir {Computing} with {Random} {Skyrmion} {Textures}},\ }\href {https://doi.org/10.1103/PhysRevApplied.14.054020} {\bibfield  {journal} {\bibinfo  {journal} {Phys. Rev. Appl.}\ }\textbf {\bibinfo {volume} {14}},\ \bibinfo {pages} {054020} (\bibinfo {year} {2020})}\BibitemShut {NoStop}%
\bibitem [{\citenamefont {Raab}\ \emph {et~al.}(2022)\citenamefont {Raab}, \citenamefont {Brems}, \citenamefont {Beneke}, \citenamefont {Dohi}, \citenamefont {Rothörl}, \citenamefont {Kammerbauer}, \citenamefont {Mentink},\ and\ \citenamefont {Kläui}}]{raab_brownian_2022}%
  \BibitemOpen
  \bibfield  {author} {\bibinfo {author} {\bibfnamefont {K.}~\bibnamefont {Raab}}, \bibinfo {author} {\bibfnamefont {M.~A.}\ \bibnamefont {Brems}}, \bibinfo {author} {\bibfnamefont {G.}~\bibnamefont {Beneke}}, \bibinfo {author} {\bibfnamefont {T.}~\bibnamefont {Dohi}}, \bibinfo {author} {\bibfnamefont {J.}~\bibnamefont {Rothörl}}, \bibinfo {author} {\bibfnamefont {F.}~\bibnamefont {Kammerbauer}}, \bibinfo {author} {\bibfnamefont {J.~H.}\ \bibnamefont {Mentink}},\ and\ \bibinfo {author} {\bibfnamefont {M.}~\bibnamefont {Kläui}},\ }\bibfield  {title} {\bibinfo {title} {Brownian reservoir computing realized using geometrically confined skyrmion dynamics},\ }\href {https://doi.org/10.1038/s41467-022-34309-2} {\bibfield  {journal} {\bibinfo  {journal} {Nat. Commun.}\ }\textbf {\bibinfo {volume} {13}},\ \bibinfo {pages} {6982} (\bibinfo {year} {2022})}\BibitemShut {NoStop}%
\bibitem [{\citenamefont {Lee}\ and\ \citenamefont {Mochizuki}(2022)}]{lee_reservoir_2022}%
  \BibitemOpen
  \bibfield  {author} {\bibinfo {author} {\bibfnamefont {M.-K.}\ \bibnamefont {Lee}}\ and\ \bibinfo {author} {\bibfnamefont {M.}~\bibnamefont {Mochizuki}},\ }\bibfield  {title} {\bibinfo {title} {Reservoir {Computing} with {Spin} {Waves} in a {Skyrmion} {Crystal}},\ }\href {https://doi.org/10.1103/PhysRevApplied.18.014074} {\bibfield  {journal} {\bibinfo  {journal} {Phys. Rev. Appl.}\ }\textbf {\bibinfo {volume} {18}},\ \bibinfo {pages} {014074} (\bibinfo {year} {2022})}\BibitemShut {NoStop}%
\bibitem [{\citenamefont {Yokouchi}\ \emph {et~al.}(2022)\citenamefont {Yokouchi}, \citenamefont {Sugimoto}, \citenamefont {Rana}, \citenamefont {Seki}, \citenamefont {Ogawa}, \citenamefont {Shiomi}, \citenamefont {Kasai},\ and\ \citenamefont {Otani}}]{yokouchi_pattern_2022}%
  \BibitemOpen
  \bibfield  {author} {\bibinfo {author} {\bibfnamefont {T.}~\bibnamefont {Yokouchi}}, \bibinfo {author} {\bibfnamefont {S.}~\bibnamefont {Sugimoto}}, \bibinfo {author} {\bibfnamefont {B.}~\bibnamefont {Rana}}, \bibinfo {author} {\bibfnamefont {S.}~\bibnamefont {Seki}}, \bibinfo {author} {\bibfnamefont {N.}~\bibnamefont {Ogawa}}, \bibinfo {author} {\bibfnamefont {Y.}~\bibnamefont {Shiomi}}, \bibinfo {author} {\bibfnamefont {S.}~\bibnamefont {Kasai}},\ and\ \bibinfo {author} {\bibfnamefont {Y.}~\bibnamefont {Otani}},\ }\bibfield  {title} {\bibinfo {title} {Pattern recognition with neuromorphic computing using magnetic field–induced dynamics of skyrmions},\ }\href {https://doi.org/10.1126/sciadv.abq5652} {\bibfield  {journal} {\bibinfo  {journal} {Sci. Adv.}\ }\textbf {\bibinfo {volume} {8}},\ \bibinfo {pages} {eabq5652} (\bibinfo {year} {2022})}\BibitemShut {NoStop}%
\bibitem [{\citenamefont {Li}\ \emph {et~al.}(2017)\citenamefont {Li}, \citenamefont {Kang}, \citenamefont {Huang}, \citenamefont {Zhang}, \citenamefont {Zhou},\ and\ \citenamefont {Zhao}}]{li_magnetic_2017}%
  \BibitemOpen
  \bibfield  {author} {\bibinfo {author} {\bibfnamefont {S.}~\bibnamefont {Li}}, \bibinfo {author} {\bibfnamefont {W.}~\bibnamefont {Kang}}, \bibinfo {author} {\bibfnamefont {Y.}~\bibnamefont {Huang}}, \bibinfo {author} {\bibfnamefont {X.}~\bibnamefont {Zhang}}, \bibinfo {author} {\bibfnamefont {Y.}~\bibnamefont {Zhou}},\ and\ \bibinfo {author} {\bibfnamefont {W.}~\bibnamefont {Zhao}},\ }\bibfield  {title} {\bibinfo {title} {Magnetic skyrmion-based artificial neuron device},\ }\href {https://doi.org/10.1088/1361-6528/aa7af5} {\bibfield  {journal} {\bibinfo  {journal} {Nanotechnology}\ }\textbf {\bibinfo {volume} {28}},\ \bibinfo {pages} {31LT01} (\bibinfo {year} {2017})}\BibitemShut {NoStop}%
\bibitem [{\citenamefont {Cai}\ and\ \citenamefont {Liu}(2021)}]{cai_current-driven_2021}%
  \BibitemOpen
  \bibfield  {author} {\bibinfo {author} {\bibfnamefont {N.}~\bibnamefont {Cai}}\ and\ \bibinfo {author} {\bibfnamefont {Y.}~\bibnamefont {Liu}},\ }\bibfield  {title} {\bibinfo {title} {Current-driven skyrmion movement in a curved nanotrack},\ }\href {https://doi.org/10.1088/1361-6463/abd12c} {\bibfield  {journal} {\bibinfo  {journal} {J. Phys. D}\ }\textbf {\bibinfo {volume} {54}},\ \bibinfo {pages} {125001} (\bibinfo {year} {2021})}\BibitemShut {NoStop}%
\bibitem [{\citenamefont {Zhang}\ \emph {et~al.}(2023)\citenamefont {Zhang}, \citenamefont {Zhang}, \citenamefont {Hou}, \citenamefont {Qin}, \citenamefont {Gao},\ and\ \citenamefont {Liu}}]{zhang_magnetic_2023}%
  \BibitemOpen
  \bibfield  {author} {\bibinfo {author} {\bibfnamefont {H.}~\bibnamefont {Zhang}}, \bibinfo {author} {\bibfnamefont {Y.}~\bibnamefont {Zhang}}, \bibinfo {author} {\bibfnamefont {Z.}~\bibnamefont {Hou}}, \bibinfo {author} {\bibfnamefont {M.}~\bibnamefont {Qin}}, \bibinfo {author} {\bibfnamefont {X.}~\bibnamefont {Gao}},\ and\ \bibinfo {author} {\bibfnamefont {J.}~\bibnamefont {Liu}},\ }\bibfield  {title} {\bibinfo {title} {Magnetic skyrmions: materials, manipulation, detection, and applications in spintronic devices},\ }\href {https://doi.org/10.1088/2752-5724/ace1df} {\bibfield  {journal} {\bibinfo  {journal} {Mater. Futures}\ }\textbf {\bibinfo {volume} {2}},\ \bibinfo {pages} {032201} (\bibinfo {year} {2023})}\BibitemShut {NoStop}%
\bibitem [{\citenamefont {Purnama}\ \emph {et~al.}(2015)\citenamefont {Purnama}, \citenamefont {Gan}, \citenamefont {Wong},\ and\ \citenamefont {Lew}}]{purnama_guided_2015}%
  \BibitemOpen
  \bibfield  {author} {\bibinfo {author} {\bibfnamefont {I.}~\bibnamefont {Purnama}}, \bibinfo {author} {\bibfnamefont {W.~L.}\ \bibnamefont {Gan}}, \bibinfo {author} {\bibfnamefont {D.~W.}\ \bibnamefont {Wong}},\ and\ \bibinfo {author} {\bibfnamefont {W.~S.}\ \bibnamefont {Lew}},\ }\bibfield  {title} {\bibinfo {title} {Guided current-induced skyrmion motion in {1D} potential well},\ }\href {https://doi.org/10.1038/srep10620} {\bibfield  {journal} {\bibinfo  {journal} {Sci. Rep.}\ }\textbf {\bibinfo {volume} {5}},\ \bibinfo {pages} {10620} (\bibinfo {year} {2015})}\BibitemShut {NoStop}%
\bibitem [{\citenamefont {Chen}\ \emph {et~al.}(2023)\citenamefont {Chen}, \citenamefont {Chen}, \citenamefont {Han}, \citenamefont {Liu}, \citenamefont {Su}, \citenamefont {Zhu}, \citenamefont {Zhou}, \citenamefont {Pan},\ and\ \citenamefont {Song}}]{chen_ordered_2023}%
  \BibitemOpen
  \bibfield  {author} {\bibinfo {author} {\bibfnamefont {R.}~\bibnamefont {Chen}}, \bibinfo {author} {\bibfnamefont {C.}~\bibnamefont {Chen}}, \bibinfo {author} {\bibfnamefont {L.}~\bibnamefont {Han}}, \bibinfo {author} {\bibfnamefont {P.}~\bibnamefont {Liu}}, \bibinfo {author} {\bibfnamefont {R.}~\bibnamefont {Su}}, \bibinfo {author} {\bibfnamefont {W.}~\bibnamefont {Zhu}}, \bibinfo {author} {\bibfnamefont {Y.}~\bibnamefont {Zhou}}, \bibinfo {author} {\bibfnamefont {F.}~\bibnamefont {Pan}},\ and\ \bibinfo {author} {\bibfnamefont {C.}~\bibnamefont {Song}},\ }\bibfield  {title} {\bibinfo {title} {Ordered creation and motion of skyrmions with surface acoustic wave},\ }\href {https://doi.org/10.1038/s41467-023-40131-1} {\bibfield  {journal} {\bibinfo  {journal} {Nat. Commun.}\ }\textbf {\bibinfo {volume} {14}},\ \bibinfo {pages} {4427} (\bibinfo {year} {2023})}\BibitemShut {NoStop}%
\bibitem [{\citenamefont {Yokouchi}\ \emph {et~al.}(2020)\citenamefont {Yokouchi}, \citenamefont {Sugimoto}, \citenamefont {Rana}, \citenamefont {Seki}, \citenamefont {Ogawa}, \citenamefont {Kasai},\ and\ \citenamefont {Otani}}]{yokouchi_creation_2020}%
  \BibitemOpen
  \bibfield  {author} {\bibinfo {author} {\bibfnamefont {T.}~\bibnamefont {Yokouchi}}, \bibinfo {author} {\bibfnamefont {S.}~\bibnamefont {Sugimoto}}, \bibinfo {author} {\bibfnamefont {B.}~\bibnamefont {Rana}}, \bibinfo {author} {\bibfnamefont {S.}~\bibnamefont {Seki}}, \bibinfo {author} {\bibfnamefont {N.}~\bibnamefont {Ogawa}}, \bibinfo {author} {\bibfnamefont {S.}~\bibnamefont {Kasai}},\ and\ \bibinfo {author} {\bibfnamefont {Y.}~\bibnamefont {Otani}},\ }\bibfield  {title} {\bibinfo {title} {Creation of magnetic skyrmions by surface acoustic waves},\ }\href {https://doi.org/10.1038/s41565-020-0661-1} {\bibfield  {journal} {\bibinfo  {journal} {Nat. Nanotechnol.}\ }\textbf {\bibinfo {volume} {15}},\ \bibinfo {pages} {361} (\bibinfo {year} {2020})}\BibitemShut {NoStop}%
\bibitem [{\citenamefont {Miyazaki}\ \emph {et~al.}(2023)\citenamefont {Miyazaki}, \citenamefont {Yokouchi},\ and\ \citenamefont {Shiomi}}]{miyazaki_trapping_2023}%
  \BibitemOpen
  \bibfield  {author} {\bibinfo {author} {\bibfnamefont {Y.}~\bibnamefont {Miyazaki}}, \bibinfo {author} {\bibfnamefont {T.}~\bibnamefont {Yokouchi}},\ and\ \bibinfo {author} {\bibfnamefont {Y.}~\bibnamefont {Shiomi}},\ }\bibfield  {title} {\bibinfo {title} {Trapping and manipulating skyrmions in two-dimensional films by surface acoustic waves},\ }\href {https://doi.org/10.1038/s41598-023-29022-z} {\bibfield  {journal} {\bibinfo  {journal} {Sci. Rep.}\ }\textbf {\bibinfo {volume} {13}},\ \bibinfo {pages} {1922} (\bibinfo {year} {2023})}\BibitemShut {NoStop}%
\bibitem [{\citenamefont {Liu}\ \emph {et~al.}(2019)\citenamefont {Liu}, \citenamefont {Huo}, \citenamefont {Xuan},\ and\ \citenamefont {Yan}}]{liu_manipulating_2019}%
  \BibitemOpen
  \bibfield  {author} {\bibinfo {author} {\bibfnamefont {Y.}~\bibnamefont {Liu}}, \bibinfo {author} {\bibfnamefont {X.}~\bibnamefont {Huo}}, \bibinfo {author} {\bibfnamefont {S.}~\bibnamefont {Xuan}},\ and\ \bibinfo {author} {\bibfnamefont {H.}~\bibnamefont {Yan}},\ }\bibfield  {title} {\bibinfo {title} {Manipulating movement of skyrmion by strain gradient in a nanotrack},\ }\href {https://doi.org/10.1016/j.jmmm.2019.165659} {\bibfield  {journal} {\bibinfo  {journal} {J. Magn. Magn. Mater}\ }\textbf {\bibinfo {volume} {492}},\ \bibinfo {pages} {165659} (\bibinfo {year} {2019})}\BibitemShut {NoStop}%
\bibitem [{\citenamefont {Yang}\ \emph {et~al.}(2024{\natexlab{a}})\citenamefont {Yang}, \citenamefont {Zhao}, \citenamefont {Yi}, \citenamefont {Xu}, \citenamefont {Chai}, \citenamefont {Zhang}, \citenamefont {Jiang}, \citenamefont {Ji}, \citenamefont {Hou}, \citenamefont {Jiang}, \citenamefont {Tang}, \citenamefont {Yu}, \citenamefont {Wu},\ and\ \citenamefont {Nan}}]{yang_acoustic-driven_2024}%
  \BibitemOpen
  \bibfield  {author} {\bibinfo {author} {\bibfnamefont {Y.}~\bibnamefont {Yang}}, \bibinfo {author} {\bibfnamefont {L.}~\bibnamefont {Zhao}}, \bibinfo {author} {\bibfnamefont {D.}~\bibnamefont {Yi}}, \bibinfo {author} {\bibfnamefont {T.}~\bibnamefont {Xu}}, \bibinfo {author} {\bibfnamefont {Y.}~\bibnamefont {Chai}}, \bibinfo {author} {\bibfnamefont {C.}~\bibnamefont {Zhang}}, \bibinfo {author} {\bibfnamefont {D.}~\bibnamefont {Jiang}}, \bibinfo {author} {\bibfnamefont {Y.}~\bibnamefont {Ji}}, \bibinfo {author} {\bibfnamefont {D.}~\bibnamefont {Hou}}, \bibinfo {author} {\bibfnamefont {W.}~\bibnamefont {Jiang}}, \bibinfo {author} {\bibfnamefont {J.}~\bibnamefont {Tang}}, \bibinfo {author} {\bibfnamefont {P.}~\bibnamefont {Yu}}, \bibinfo {author} {\bibfnamefont {H.}~\bibnamefont {Wu}},\ and\ \bibinfo {author} {\bibfnamefont {T.}~\bibnamefont {Nan}},\ }\bibfield  {title} {\bibinfo {title} {Acoustic-driven magnetic skyrmion motion},\ }\href {https://doi.org/10.1038/s41467-024-45316-w} {\bibfield  {journal}
  {\bibinfo  {journal} {Nat. Commun.}\ }\textbf {\bibinfo {volume} {15}},\ \bibinfo {pages} {1018} (\bibinfo {year} {2024}{\natexlab{a}})}\BibitemShut {NoStop}%
\bibitem [{\citenamefont {Shuai}\ \emph {et~al.}(2024)\citenamefont {Shuai}, \citenamefont {Lopez-Diaz}, \citenamefont {Cunningham},\ and\ \citenamefont {Moore}}]{shuai_transport_2024}%
  \BibitemOpen
  \bibfield  {author} {\bibinfo {author} {\bibfnamefont {J.}~\bibnamefont {Shuai}}, \bibinfo {author} {\bibfnamefont {L.}~\bibnamefont {Lopez-Diaz}}, \bibinfo {author} {\bibfnamefont {J.~E.}\ \bibnamefont {Cunningham}},\ and\ \bibinfo {author} {\bibfnamefont {T.~A.}\ \bibnamefont {Moore}},\ }\bibfield  {title} {\bibinfo {title} {Transport of skyrmions by surface acoustic waves},\ }\href {https://doi.org/10.1063/5.0207929} {\bibfield  {journal} {\bibinfo  {journal} {Appl. Phys. Lett.}\ }\textbf {\bibinfo {volume} {124}},\ \bibinfo {pages} {202407} (\bibinfo {year} {2024})}\BibitemShut {NoStop}%
\bibitem [{\citenamefont {Koujok}\ \emph {et~al.}(2023)\citenamefont {Koujok}, \citenamefont {Riveros}, \citenamefont {Rodrigues}, \citenamefont {Finocchio}, \citenamefont {Weiler}, \citenamefont {Hamadeh},\ and\ \citenamefont {Pirro}}]{koujok_resonant_2023}%
  \BibitemOpen
  \bibfield  {author} {\bibinfo {author} {\bibfnamefont {A.}~\bibnamefont {Koujok}}, \bibinfo {author} {\bibfnamefont {A.}~\bibnamefont {Riveros}}, \bibinfo {author} {\bibfnamefont {D.~R.}\ \bibnamefont {Rodrigues}}, \bibinfo {author} {\bibfnamefont {G.}~\bibnamefont {Finocchio}}, \bibinfo {author} {\bibfnamefont {M.}~\bibnamefont {Weiler}}, \bibinfo {author} {\bibfnamefont {A.}~\bibnamefont {Hamadeh}},\ and\ \bibinfo {author} {\bibfnamefont {P.}~\bibnamefont {Pirro}},\ }\bibfield  {title} {\bibinfo {title} {Resonant excitation of vortex gyrotropic mode via surface acoustic waves},\ }\href {https://doi.org/10.1063/5.0168968} {\bibfield  {journal} {\bibinfo  {journal} {Appl. Phys. Lett.}\ }\textbf {\bibinfo {volume} {123}},\ \bibinfo {pages} {132403} (\bibinfo {year} {2023})}\BibitemShut {NoStop}%
\bibitem [{\citenamefont {Iurchuk}\ \emph {et~al.}(2024)\citenamefont {Iurchuk}, \citenamefont {Lindner}, \citenamefont {Fassbender},\ and\ \citenamefont {Kákay}}]{iurchuk_excitation_2024}%
  \BibitemOpen
  \bibfield  {author} {\bibinfo {author} {\bibfnamefont {V.}~\bibnamefont {Iurchuk}}, \bibinfo {author} {\bibfnamefont {J.}~\bibnamefont {Lindner}}, \bibinfo {author} {\bibfnamefont {J.}~\bibnamefont {Fassbender}},\ and\ \bibinfo {author} {\bibfnamefont {A.}~\bibnamefont {Kákay}},\ }\bibfield  {title} {\bibinfo {title} {Excitation of the {Gyrotropic} {Mode} in a {Magnetic} {Vortex} by {Time}-{Varying} {Strain}},\ }\href {https://doi.org/10.1103/PhysRevLett.133.146701} {\bibfield  {journal} {\bibinfo  {journal} {Phys. Rev. Lett.}\ }\textbf {\bibinfo {volume} {133}},\ \bibinfo {pages} {146701} (\bibinfo {year} {2024})}\BibitemShut {NoStop}%
\bibitem [{\citenamefont {Moukhader}\ \emph {et~al.}(2024)\citenamefont {Moukhader}, \citenamefont {Rodrigues}, \citenamefont {Riveros}, \citenamefont {Koujok}, \citenamefont {Finocchio}, \citenamefont {Pirro},\ and\ \citenamefont {Hamadeh}}]{moukhader_injection_2024}%
  \BibitemOpen
  \bibfield  {author} {\bibinfo {author} {\bibfnamefont {R.}~\bibnamefont {Moukhader}}, \bibinfo {author} {\bibfnamefont {D.~R.}\ \bibnamefont {Rodrigues}}, \bibinfo {author} {\bibfnamefont {A.}~\bibnamefont {Riveros}}, \bibinfo {author} {\bibfnamefont {A.}~\bibnamefont {Koujok}}, \bibinfo {author} {\bibfnamefont {G.}~\bibnamefont {Finocchio}}, \bibinfo {author} {\bibfnamefont {P.}~\bibnamefont {Pirro}},\ and\ \bibinfo {author} {\bibfnamefont {A.}~\bibnamefont {Hamadeh}},\ }\bibfield  {title} {\bibinfo {title} {Injection locking in {DC}-driven spintronic vortex oscillators via surface acoustic wave modulation},\ }\href {https://doi.org/10.1063/5.0225582} {\bibfield  {journal} {\bibinfo  {journal} {J. Appl. Phys.}\ }\textbf {\bibinfo {volume} {136}},\ \bibinfo {pages} {183901} (\bibinfo {year} {2024})}\BibitemShut {NoStop}%
\bibitem [{\citenamefont {Seeger}\ \emph {et~al.}(2024)\citenamefont {Seeger}, \citenamefont {Millo}, \citenamefont {Soares}, \citenamefont {Kim}, \citenamefont {Solignac}, \citenamefont {Loubens},\ and\ \citenamefont {Devolder}}]{seeger_experimental_2024}%
  \BibitemOpen
  \bibfield  {author} {\bibinfo {author} {\bibfnamefont {R.~L.}\ \bibnamefont {Seeger}}, \bibinfo {author} {\bibfnamefont {F.}~\bibnamefont {Millo}}, \bibinfo {author} {\bibfnamefont {G.}~\bibnamefont {Soares}}, \bibinfo {author} {\bibfnamefont {J.-V.}\ \bibnamefont {Kim}}, \bibinfo {author} {\bibfnamefont {A.}~\bibnamefont {Solignac}}, \bibinfo {author} {\bibfnamefont {G.~d.}\ \bibnamefont {Loubens}},\ and\ \bibinfo {author} {\bibfnamefont {T.}~\bibnamefont {Devolder}},\ }\bibfield  {title} {\bibinfo {title} {Experimental observation of vortex gyrotropic mode excited by surface acoustic waves},\ }\bibfield  {journal} {\bibinfo  {journal} {arXiv}\ }\href {https://doi.org/10.48550/arXiv.2409.05998} {10.48550/arXiv.2409.05998} (\bibinfo {year} {2024})\BibitemShut {NoStop}%
\bibitem [{\citenamefont {Gong}\ \emph {et~al.}(2022)\citenamefont {Gong}, \citenamefont {Jing}, \citenamefont {Lu},\ and\ \citenamefont {Wang}}]{gong_skyrmion_2022}%
  \BibitemOpen
  \bibfield  {author} {\bibinfo {author} {\bibfnamefont {X.}~\bibnamefont {Gong}}, \bibinfo {author} {\bibfnamefont {K.~Y.}\ \bibnamefont {Jing}}, \bibinfo {author} {\bibfnamefont {J.}~\bibnamefont {Lu}},\ and\ \bibinfo {author} {\bibfnamefont {X.~R.}\ \bibnamefont {Wang}},\ }\bibfield  {title} {\bibinfo {title} {Skyrmion pinning by disk-shaped defects},\ }\href {https://doi.org/10.1103/PhysRevB.105.094437} {\bibfield  {journal} {\bibinfo  {journal} {Phys. Rev. B}\ }\textbf {\bibinfo {volume} {105}},\ \bibinfo {pages} {094437} (\bibinfo {year} {2022})}\BibitemShut {NoStop}%
\bibitem [{\citenamefont {Gruber}\ \emph {et~al.}(2022)\citenamefont {Gruber}, \citenamefont {Zázvorka}, \citenamefont {Brems}, \citenamefont {Rodrigues}, \citenamefont {Dohi}, \citenamefont {Kerber}, \citenamefont {Seng}, \citenamefont {Vafaee}, \citenamefont {Everschor-Sitte}, \citenamefont {Virnau},\ and\ \citenamefont {Kläui}}]{gruber_skyrmion_2022}%
  \BibitemOpen
  \bibfield  {author} {\bibinfo {author} {\bibfnamefont {R.}~\bibnamefont {Gruber}}, \bibinfo {author} {\bibfnamefont {J.}~\bibnamefont {Zázvorka}}, \bibinfo {author} {\bibfnamefont {M.~A.}\ \bibnamefont {Brems}}, \bibinfo {author} {\bibfnamefont {D.~R.}\ \bibnamefont {Rodrigues}}, \bibinfo {author} {\bibfnamefont {T.}~\bibnamefont {Dohi}}, \bibinfo {author} {\bibfnamefont {N.}~\bibnamefont {Kerber}}, \bibinfo {author} {\bibfnamefont {B.}~\bibnamefont {Seng}}, \bibinfo {author} {\bibfnamefont {M.}~\bibnamefont {Vafaee}}, \bibinfo {author} {\bibfnamefont {K.}~\bibnamefont {Everschor-Sitte}}, \bibinfo {author} {\bibfnamefont {P.}~\bibnamefont {Virnau}},\ and\ \bibinfo {author} {\bibfnamefont {M.}~\bibnamefont {Kläui}},\ }\bibfield  {title} {\bibinfo {title} {Skyrmion pinning energetics in thin film systems},\ }\href {https://doi.org/10.1038/s41467-022-30743-4} {\bibfield  {journal} {\bibinfo  {journal} {Nat. Commun.}\ }\textbf {\bibinfo {volume} {13}},\ \bibinfo {pages} {3144} (\bibinfo {year}
  {2022})}\BibitemShut {NoStop}%
\bibitem [{\citenamefont {Stosic}\ \emph {et~al.}(2017)\citenamefont {Stosic}, \citenamefont {Ludermir},\ and\ \citenamefont {Milošević}}]{stosic_pinning_2017}%
  \BibitemOpen
  \bibfield  {author} {\bibinfo {author} {\bibfnamefont {D.}~\bibnamefont {Stosic}}, \bibinfo {author} {\bibfnamefont {T.~B.}\ \bibnamefont {Ludermir}},\ and\ \bibinfo {author} {\bibfnamefont {M.~V.}\ \bibnamefont {Milošević}},\ }\bibfield  {title} {\bibinfo {title} {Pinning of magnetic skyrmions in a monolayer {Co} film on {Pt}(111): {Theoretical} characterization and exemplified utilization},\ }\href {https://doi.org/10.1103/PhysRevB.96.214403} {\bibfield  {journal} {\bibinfo  {journal} {Phys. Rev. B}\ }\textbf {\bibinfo {volume} {96}},\ \bibinfo {pages} {214403} (\bibinfo {year} {2017})}\BibitemShut {NoStop}%
\bibitem [{\citenamefont {Schülein}\ \emph {et~al.}(2015)\citenamefont {Schülein}, \citenamefont {Zallo}, \citenamefont {Atkinson}, \citenamefont {Schmidt}, \citenamefont {Trotta}, \citenamefont {Rastelli}, \citenamefont {Wixforth},\ and\ \citenamefont {Krenner}}]{schulein_fourier_2015}%
  \BibitemOpen
  \bibfield  {author} {\bibinfo {author} {\bibfnamefont {F.~J.~R.}\ \bibnamefont {Schülein}}, \bibinfo {author} {\bibfnamefont {E.}~\bibnamefont {Zallo}}, \bibinfo {author} {\bibfnamefont {P.}~\bibnamefont {Atkinson}}, \bibinfo {author} {\bibfnamefont {O.~G.}\ \bibnamefont {Schmidt}}, \bibinfo {author} {\bibfnamefont {R.}~\bibnamefont {Trotta}}, \bibinfo {author} {\bibfnamefont {A.}~\bibnamefont {Rastelli}}, \bibinfo {author} {\bibfnamefont {A.}~\bibnamefont {Wixforth}},\ and\ \bibinfo {author} {\bibfnamefont {H.~J.}\ \bibnamefont {Krenner}},\ }\bibfield  {title} {\bibinfo {title} {Fourier synthesis of radiofrequency nanomechanical pulses with different shapes},\ }\href {https://doi.org/10.1038/nnano.2015.72} {\bibfield  {journal} {\bibinfo  {journal} {Nat. Nanotechnol.}\ }\textbf {\bibinfo {volume} {10}},\ \bibinfo {pages} {512} (\bibinfo {year} {2015})}\BibitemShut {NoStop}%
\bibitem [{\citenamefont {Weiß}\ \emph {et~al.}(2018)\citenamefont {Weiß}, \citenamefont {Hörner}, \citenamefont {Zallo}, \citenamefont {Atkinson}, \citenamefont {Rastelli}, \citenamefont {Schmidt}, \citenamefont {Wixforth},\ and\ \citenamefont {Krenner}}]{weis_multiharmonic_2018}%
  \BibitemOpen
  \bibfield  {author} {\bibinfo {author} {\bibfnamefont {M.}~\bibnamefont {Weiß}}, \bibinfo {author} {\bibfnamefont {A.~L.}\ \bibnamefont {Hörner}}, \bibinfo {author} {\bibfnamefont {E.}~\bibnamefont {Zallo}}, \bibinfo {author} {\bibfnamefont {P.}~\bibnamefont {Atkinson}}, \bibinfo {author} {\bibfnamefont {A.}~\bibnamefont {Rastelli}}, \bibinfo {author} {\bibfnamefont {O.~G.}\ \bibnamefont {Schmidt}}, \bibinfo {author} {\bibfnamefont {A.}~\bibnamefont {Wixforth}},\ and\ \bibinfo {author} {\bibfnamefont {H.~J.}\ \bibnamefont {Krenner}},\ }\bibfield  {title} {\bibinfo {title} {Multiharmonic {Frequency}-{Chirped} {Transducers} for {Surface}-{Acoustic}-{Wave} {Optomechanics}},\ }\href {https://doi.org/10.1103/PhysRevApplied.9.014004} {\bibfield  {journal} {\bibinfo  {journal} {Phys. Rev. Appl}\ }\textbf {\bibinfo {volume} {9}},\ \bibinfo {pages} {014004} (\bibinfo {year} {2018})}\BibitemShut {NoStop}%
\bibitem [{\citenamefont {Wang}\ \emph {et~al.}(2022)\citenamefont {Wang}, \citenamefont {Song}, \citenamefont {Wei}, \citenamefont {Nan}, \citenamefont {Zhang}, \citenamefont {Ge}, \citenamefont {Tian}, \citenamefont {Zang},\ and\ \citenamefont {Du}}]{wang_electrical_2022}%
  \BibitemOpen
  \bibfield  {author} {\bibinfo {author} {\bibfnamefont {W.}~\bibnamefont {Wang}}, \bibinfo {author} {\bibfnamefont {D.}~\bibnamefont {Song}}, \bibinfo {author} {\bibfnamefont {W.}~\bibnamefont {Wei}}, \bibinfo {author} {\bibfnamefont {P.}~\bibnamefont {Nan}}, \bibinfo {author} {\bibfnamefont {S.}~\bibnamefont {Zhang}}, \bibinfo {author} {\bibfnamefont {B.}~\bibnamefont {Ge}}, \bibinfo {author} {\bibfnamefont {M.}~\bibnamefont {Tian}}, \bibinfo {author} {\bibfnamefont {J.}~\bibnamefont {Zang}},\ and\ \bibinfo {author} {\bibfnamefont {H.}~\bibnamefont {Du}},\ }\bibfield  {title} {\bibinfo {title} {Electrical manipulation of skyrmions in a chiral magnet},\ }\href {https://doi.org/10.1038/s41467-022-29217-4} {\bibfield  {journal} {\bibinfo  {journal} {Nat. Commun.}\ }\textbf {\bibinfo {volume} {13}},\ \bibinfo {pages} {1593} (\bibinfo {year} {2022})}\BibitemShut {NoStop}%
\bibitem [{\citenamefont {De~Jong}\ \emph {et~al.}(2023)\citenamefont {De~Jong}, \citenamefont {Smit}, \citenamefont {Meijer}, \citenamefont {Lucassen}, \citenamefont {Swagten}, \citenamefont {Koopmans},\ and\ \citenamefont {Lavrijsen}}]{de_jong_controlling_2023}%
  \BibitemOpen
  \bibfield  {author} {\bibinfo {author} {\bibfnamefont {M.~C.~H.}\ \bibnamefont {De~Jong}}, \bibinfo {author} {\bibfnamefont {B.~H.~M.}\ \bibnamefont {Smit}}, \bibinfo {author} {\bibfnamefont {M.~J.}\ \bibnamefont {Meijer}}, \bibinfo {author} {\bibfnamefont {J.}~\bibnamefont {Lucassen}}, \bibinfo {author} {\bibfnamefont {H.~J.~M.}\ \bibnamefont {Swagten}}, \bibinfo {author} {\bibfnamefont {B.}~\bibnamefont {Koopmans}},\ and\ \bibinfo {author} {\bibfnamefont {R.}~\bibnamefont {Lavrijsen}},\ }\bibfield  {title} {\bibinfo {title} {Controlling magnetic skyrmion nucleation with {Ga} + ion irradiation},\ }\href {https://doi.org/10.1103/PhysRevB.107.094429} {\bibfield  {journal} {\bibinfo  {journal} {Phys. Rev. B}\ }\textbf {\bibinfo {volume} {107}},\ \bibinfo {pages} {094429} (\bibinfo {year} {2023})}\BibitemShut {NoStop}%
\bibitem [{\citenamefont {Fallon}\ \emph {et~al.}(2020)\citenamefont {Fallon}, \citenamefont {Hughes}, \citenamefont {Zeissler}, \citenamefont {Legrand}, \citenamefont {Ajejas}, \citenamefont {Maccariello}, \citenamefont {McFadzean}, \citenamefont {Smith}, \citenamefont {McGrouther}, \citenamefont {Collin}, \citenamefont {Reyren}, \citenamefont {Cros}, \citenamefont {Marrows},\ and\ \citenamefont {McVitie}}]{fallon_controlled_2020}%
  \BibitemOpen
  \bibfield  {author} {\bibinfo {author} {\bibfnamefont {K.}~\bibnamefont {Fallon}}, \bibinfo {author} {\bibfnamefont {S.}~\bibnamefont {Hughes}}, \bibinfo {author} {\bibfnamefont {K.}~\bibnamefont {Zeissler}}, \bibinfo {author} {\bibfnamefont {W.}~\bibnamefont {Legrand}}, \bibinfo {author} {\bibfnamefont {F.}~\bibnamefont {Ajejas}}, \bibinfo {author} {\bibfnamefont {D.}~\bibnamefont {Maccariello}}, \bibinfo {author} {\bibfnamefont {S.}~\bibnamefont {McFadzean}}, \bibinfo {author} {\bibfnamefont {W.}~\bibnamefont {Smith}}, \bibinfo {author} {\bibfnamefont {D.}~\bibnamefont {McGrouther}}, \bibinfo {author} {\bibfnamefont {S.}~\bibnamefont {Collin}}, \bibinfo {author} {\bibfnamefont {N.}~\bibnamefont {Reyren}}, \bibinfo {author} {\bibfnamefont {V.}~\bibnamefont {Cros}}, \bibinfo {author} {\bibfnamefont {C.~H.}\ \bibnamefont {Marrows}},\ and\ \bibinfo {author} {\bibfnamefont {S.}~\bibnamefont {McVitie}},\ }\bibfield  {title} {\bibinfo {title} {Controlled {Individual} {Skyrmion} {Nucleation} at {Artificial}
  {Defects} {Formed} by {Ion} {Irradiation}},\ }\href {https://doi.org/10.1002/smll.201907450} {\bibfield  {journal} {\bibinfo  {journal} {Small}\ }\textbf {\bibinfo {volume} {16}},\ \bibinfo {pages} {1907450} (\bibinfo {year} {2020})}\BibitemShut {NoStop}%
\bibitem [{\citenamefont {Zhao}\ \emph {et~al.}(2023)\citenamefont {Zhao}, \citenamefont {Wang}, \citenamefont {Xu}, \citenamefont {Yu}, \citenamefont {Hou}, \citenamefont {Meng}, \citenamefont {Xie}, \citenamefont {Meng}, \citenamefont {Zhu}, \citenamefont {Hou}, \citenamefont {Yang}, \citenamefont {Luo}, \citenamefont {Wu}, \citenamefont {Xu}, \citenamefont {Gao}, \citenamefont {Feng},\ and\ \citenamefont {Yu}}]{zhao_local_2023}%
  \BibitemOpen
  \bibfield  {author} {\bibinfo {author} {\bibfnamefont {Y.}~\bibnamefont {Zhao}}, \bibinfo {author} {\bibfnamefont {J.}~\bibnamefont {Wang}}, \bibinfo {author} {\bibfnamefont {L.}~\bibnamefont {Xu}}, \bibinfo {author} {\bibfnamefont {P.}~\bibnamefont {Yu}}, \bibinfo {author} {\bibfnamefont {M.}~\bibnamefont {Hou}}, \bibinfo {author} {\bibfnamefont {F.}~\bibnamefont {Meng}}, \bibinfo {author} {\bibfnamefont {S.}~\bibnamefont {Xie}}, \bibinfo {author} {\bibfnamefont {Y.}~\bibnamefont {Meng}}, \bibinfo {author} {\bibfnamefont {R.}~\bibnamefont {Zhu}}, \bibinfo {author} {\bibfnamefont {Z.}~\bibnamefont {Hou}}, \bibinfo {author} {\bibfnamefont {M.}~\bibnamefont {Yang}}, \bibinfo {author} {\bibfnamefont {J.}~\bibnamefont {Luo}}, \bibinfo {author} {\bibfnamefont {J.}~\bibnamefont {Wu}}, \bibinfo {author} {\bibfnamefont {Y.}~\bibnamefont {Xu}}, \bibinfo {author} {\bibfnamefont {X.}~\bibnamefont {Gao}}, \bibinfo {author} {\bibfnamefont {C.}~\bibnamefont {Feng}},\ and\ \bibinfo {author} {\bibfnamefont
  {G.}~\bibnamefont {Yu}},\ }\bibfield  {title} {\bibinfo {title} {Local {Manipulation} of {Skyrmion} {Nucleation} in {Microscale} {Areas} of a {Thin} {Film} with {Nitrogen}-{Ion} {Implantation}},\ }\href {https://doi.org/10.1021/acsami.3c00266} {\bibfield  {journal} {\bibinfo  {journal} {ACS Appl. Mate.r Interfaces}\ }\textbf {\bibinfo {volume} {15}},\ \bibinfo {pages} {15004} (\bibinfo {year} {2023})}\BibitemShut {NoStop}%
\bibitem [{\citenamefont {Yu}\ \emph {et~al.}(2017)\citenamefont {Yu}, \citenamefont {Morikawa}, \citenamefont {Tokunaga}, \citenamefont {Kubota}, \citenamefont {Kurumaji}, \citenamefont {Oike}, \citenamefont {Nakamura}, \citenamefont {Kagawa}, \citenamefont {Taguchi}, \citenamefont {Arima}, \citenamefont {Kawasaki},\ and\ \citenamefont {Tokura}}]{yu_current-induced_2017}%
  \BibitemOpen
  \bibfield  {author} {\bibinfo {author} {\bibfnamefont {X.}~\bibnamefont {Yu}}, \bibinfo {author} {\bibfnamefont {D.}~\bibnamefont {Morikawa}}, \bibinfo {author} {\bibfnamefont {Y.}~\bibnamefont {Tokunaga}}, \bibinfo {author} {\bibfnamefont {M.}~\bibnamefont {Kubota}}, \bibinfo {author} {\bibfnamefont {T.}~\bibnamefont {Kurumaji}}, \bibinfo {author} {\bibfnamefont {H.}~\bibnamefont {Oike}}, \bibinfo {author} {\bibfnamefont {M.}~\bibnamefont {Nakamura}}, \bibinfo {author} {\bibfnamefont {F.}~\bibnamefont {Kagawa}}, \bibinfo {author} {\bibfnamefont {Y.}~\bibnamefont {Taguchi}}, \bibinfo {author} {\bibfnamefont {T.-h.}\ \bibnamefont {Arima}}, \bibinfo {author} {\bibfnamefont {M.}~\bibnamefont {Kawasaki}},\ and\ \bibinfo {author} {\bibfnamefont {Y.}~\bibnamefont {Tokura}},\ }\bibfield  {title} {\bibinfo {title} {Current-{Induced} {Nucleation} and {Annihilation} of {Magnetic} {Skyrmions} at {Room} {Temperature} in a {Chiral} {Magnet}},\ }\href {https://doi.org/10.1002/adma.201606178} {\bibfield  {journal}
  {\bibinfo  {journal} {Adv. Mater.}\ }\textbf {\bibinfo {volume} {29}},\ \bibinfo {pages} {1606178} (\bibinfo {year} {2017})}\BibitemShut {NoStop}%
\bibitem [{\citenamefont {Akhtar}\ \emph {et~al.}(2019)\citenamefont {Akhtar}, \citenamefont {Hrabec}, \citenamefont {Chouaieb}, \citenamefont {Haykal}, \citenamefont {Gross}, \citenamefont {Belmeguenai}, \citenamefont {Gabor}, \citenamefont {Shields}, \citenamefont {Maletinsky}, \citenamefont {Thiaville}, \citenamefont {Rohart},\ and\ \citenamefont {Jacques}}]{akhtar_current-induced_2019}%
  \BibitemOpen
  \bibfield  {author} {\bibinfo {author} {\bibfnamefont {W.}~\bibnamefont {Akhtar}}, \bibinfo {author} {\bibfnamefont {A.}~\bibnamefont {Hrabec}}, \bibinfo {author} {\bibfnamefont {S.}~\bibnamefont {Chouaieb}}, \bibinfo {author} {\bibfnamefont {A.}~\bibnamefont {Haykal}}, \bibinfo {author} {\bibfnamefont {I.}~\bibnamefont {Gross}}, \bibinfo {author} {\bibfnamefont {M.}~\bibnamefont {Belmeguenai}}, \bibinfo {author} {\bibfnamefont {M.}~\bibnamefont {Gabor}}, \bibinfo {author} {\bibfnamefont {B.}~\bibnamefont {Shields}}, \bibinfo {author} {\bibfnamefont {P.}~\bibnamefont {Maletinsky}}, \bibinfo {author} {\bibfnamefont {A.}~\bibnamefont {Thiaville}}, \bibinfo {author} {\bibfnamefont {S.}~\bibnamefont {Rohart}},\ and\ \bibinfo {author} {\bibfnamefont {V.}~\bibnamefont {Jacques}},\ }\bibfield  {title} {\bibinfo {title} {Current-{Induced} {Nucleation} and {Dynamics} of {Skyrmions} in a {Co} -based {Heusler} {Alloy}},\ }\href {https://doi.org/10.1103/PhysRevApplied.11.034066} {\bibfield  {journal} {\bibinfo
  {journal} {Phys. Rev. Appl.}\ }\textbf {\bibinfo {volume} {11}},\ \bibinfo {pages} {034066} (\bibinfo {year} {2019})}\BibitemShut {NoStop}%
\bibitem [{\citenamefont {Mallick}\ \emph {et~al.}(2022)\citenamefont {Mallick}, \citenamefont {Panigrahy}, \citenamefont {Pradhan},\ and\ \citenamefont {Rohart}}]{mallick_current-induced_2022}%
  \BibitemOpen
  \bibfield  {author} {\bibinfo {author} {\bibfnamefont {S.}~\bibnamefont {Mallick}}, \bibinfo {author} {\bibfnamefont {S.}~\bibnamefont {Panigrahy}}, \bibinfo {author} {\bibfnamefont {G.}~\bibnamefont {Pradhan}},\ and\ \bibinfo {author} {\bibfnamefont {S.}~\bibnamefont {Rohart}},\ }\bibfield  {title} {\bibinfo {title} {Current-{Induced} {Nucleation} and {Motion} of {Skyrmions} in {Zero} {Magnetic} {Field}},\ }\href {https://doi.org/10.1103/PhysRevApplied.18.064072} {\bibfield  {journal} {\bibinfo  {journal} {Phys. Rev. Appl.}\ }\textbf {\bibinfo {volume} {18}},\ \bibinfo {pages} {064072} (\bibinfo {year} {2022})}\BibitemShut {NoStop}%
\bibitem [{\citenamefont {Quessab}\ \emph {et~al.}(2022)\citenamefont {Quessab}, \citenamefont {Xu}, \citenamefont {Cogulu}, \citenamefont {Finizio}, \citenamefont {Raabe},\ and\ \citenamefont {Kent}}]{quessab_zero-field_2022}%
  \BibitemOpen
  \bibfield  {author} {\bibinfo {author} {\bibfnamefont {Y.}~\bibnamefont {Quessab}}, \bibinfo {author} {\bibfnamefont {J.-W.}\ \bibnamefont {Xu}}, \bibinfo {author} {\bibfnamefont {E.}~\bibnamefont {Cogulu}}, \bibinfo {author} {\bibfnamefont {S.}~\bibnamefont {Finizio}}, \bibinfo {author} {\bibfnamefont {J.}~\bibnamefont {Raabe}},\ and\ \bibinfo {author} {\bibfnamefont {A.~D.}\ \bibnamefont {Kent}},\ }\bibfield  {title} {\bibinfo {title} {Zero-{Field} {Nucleation} and {Fast} {Motion} of {Skyrmions} {Induced} by {Nanosecond} {Current} {Pulses} in a {Ferrimagnetic} {Thin} {Film}},\ }\href {https://doi.org/10.1021/acs.nanolett.2c01038} {\bibfield  {journal} {\bibinfo  {journal} {Nano Lett.}\ }\textbf {\bibinfo {volume} {22}},\ \bibinfo {pages} {6091} (\bibinfo {year} {2022})}\BibitemShut {NoStop}%
\bibitem [{\citenamefont {Hrabec}\ \emph {et~al.}(2017)\citenamefont {Hrabec}, \citenamefont {Sampaio}, \citenamefont {Belmeguenai}, \citenamefont {Gross}, \citenamefont {Weil}, \citenamefont {Chérif}, \citenamefont {Stashkevich}, \citenamefont {Jacques}, \citenamefont {Thiaville},\ and\ \citenamefont {Rohart}}]{hrabec_current-induced_2017}%
  \BibitemOpen
  \bibfield  {author} {\bibinfo {author} {\bibfnamefont {A.}~\bibnamefont {Hrabec}}, \bibinfo {author} {\bibfnamefont {J.}~\bibnamefont {Sampaio}}, \bibinfo {author} {\bibfnamefont {M.}~\bibnamefont {Belmeguenai}}, \bibinfo {author} {\bibfnamefont {I.}~\bibnamefont {Gross}}, \bibinfo {author} {\bibfnamefont {R.}~\bibnamefont {Weil}}, \bibinfo {author} {\bibfnamefont {S.~M.}\ \bibnamefont {Chérif}}, \bibinfo {author} {\bibfnamefont {A.}~\bibnamefont {Stashkevich}}, \bibinfo {author} {\bibfnamefont {V.}~\bibnamefont {Jacques}}, \bibinfo {author} {\bibfnamefont {A.}~\bibnamefont {Thiaville}},\ and\ \bibinfo {author} {\bibfnamefont {S.}~\bibnamefont {Rohart}},\ }\bibfield  {title} {\bibinfo {title} {Current-induced skyrmion generation and dynamics in symmetric bilayers},\ }\href {https://doi.org/10.1038/ncomms15765} {\bibfield  {journal} {\bibinfo  {journal} {Nat. Commun.}\ }\textbf {\bibinfo {volume} {8}},\ \bibinfo {pages} {15765} (\bibinfo {year} {2017})}\BibitemShut {NoStop}%
\bibitem [{\citenamefont {Finizio}\ \emph {et~al.}(2019)\citenamefont {Finizio}, \citenamefont {Zeissler}, \citenamefont {Wintz}, \citenamefont {Mayr}, \citenamefont {Weßels}, \citenamefont {Huxtable}, \citenamefont {Burnell}, \citenamefont {Marrows},\ and\ \citenamefont {Raabe}}]{finizio_deterministic_2019}%
  \BibitemOpen
  \bibfield  {author} {\bibinfo {author} {\bibfnamefont {S.}~\bibnamefont {Finizio}}, \bibinfo {author} {\bibfnamefont {K.}~\bibnamefont {Zeissler}}, \bibinfo {author} {\bibfnamefont {S.}~\bibnamefont {Wintz}}, \bibinfo {author} {\bibfnamefont {S.}~\bibnamefont {Mayr}}, \bibinfo {author} {\bibfnamefont {T.}~\bibnamefont {Weßels}}, \bibinfo {author} {\bibfnamefont {A.~J.}\ \bibnamefont {Huxtable}}, \bibinfo {author} {\bibfnamefont {G.}~\bibnamefont {Burnell}}, \bibinfo {author} {\bibfnamefont {C.~H.}\ \bibnamefont {Marrows}},\ and\ \bibinfo {author} {\bibfnamefont {J.}~\bibnamefont {Raabe}},\ }\bibfield  {title} {\bibinfo {title} {Deterministic {Field}-{Free} {Skyrmion} {Nucleation} at a {Nanoengineered} {Injector} {Device}},\ }\href {https://doi.org/10.1021/acs.nanolett.9b02840} {\bibfield  {journal} {\bibinfo  {journal} {Nano Lett.}\ }\textbf {\bibinfo {volume} {19}},\ \bibinfo {pages} {7246} (\bibinfo {year} {2019})}\BibitemShut {NoStop}%
\bibitem [{\citenamefont {Rohart}(2013)}]{rohart_skyrmion_2013}%
  \BibitemOpen
  \bibfield  {author} {\bibinfo {author} {\bibfnamefont {S.}~\bibnamefont {Rohart}},\ and\ \bibinfo {author} {\bibfnamefont {A.}~\bibnamefont {Thiaville}},\ } \bibfield  {title} {\bibinfo {title} {Skyrmion confinement in ultrathin film nanostructures in the presence of Dzyaloshinskii-Moriya interaction},\ }\href {https://doi.org/10.1103/PhysRevB.88.184422}{\bibfield  {journal} {\bibinfo  {journal} {Phys. Rev. B}\ }\textbf {\bibinfo {volume} {88}},\ (\bibinfo {year} {2013})}\BibitemShut {NoStop}%
\bibitem [{\citenamefont {Wang}\ \citenamefont {Yuan}\ and\ \citenamefont {Wang}(2018)}]{wang_theory_2018}%
  \BibitemOpen
  \bibfield  {author} {\bibinfo {author} {\bibfnamefont {X.~S.}\ \bibnamefont {Wang}},\ 
  \bibinfo {author} {\bibfnamefont {H.~Y.}\ \bibnamefont {Yuan}},\ and\ 
  \bibinfo {author} {\bibfnamefont {X.~R.}\ \bibnamefont {Wang}},\ }%
  \bibfield  {title} {\bibinfo {title} {A theory on skyrmion size},\ }%
  \href {https://doi.org/10.1038/s42005-018-0029-0} {\bibfield  {journal} {\bibinfo  {journal} {Commun. Phys.}\ }\textbf {\bibinfo {volume} {1}},\ \bibinfo {pages} {31} (\bibinfo {year} {2018})}%
  \BibitemShut {NoStop}%
\bibitem [{\citenamefont {Dreher}\ \emph {et~al.}(2012)\citenamefont {Dreher}, \citenamefont {Weiler}, \citenamefont {Pernpeintner}, \citenamefont {Huebl}, \citenamefont {Gross}, \citenamefont {Brandt},\ and\ \citenamefont {Goennenwein}}]{dreher_surface_2012}%
  \BibitemOpen
  \bibfield  {author} {\bibinfo {author} {\bibfnamefont {L.}~\bibnamefont {Dreher}}, \bibinfo {author} {\bibfnamefont {M.}~\bibnamefont {Weiler}}, \bibinfo {author} {\bibfnamefont {M.}~\bibnamefont {Pernpeintner}}, \bibinfo {author} {\bibfnamefont {H.}~\bibnamefont {Huebl}}, \bibinfo {author} {\bibfnamefont {R.}~\bibnamefont {Gross}}, \bibinfo {author} {\bibfnamefont {M.~S.}\ \bibnamefont {Brandt}},\ and\ \bibinfo {author} {\bibfnamefont {S.~T.~B.}\ \bibnamefont {Goennenwein}},\ }\bibfield  {title} {\bibinfo {title} {Surface acoustic wave driven ferromagnetic resonance in nickel thin films: {Theory} and experiment},\ }\href {https://doi.org/10.1103/PhysRevB.86.134415} {\bibfield  {journal} {\bibinfo  {journal} {Phys. Rev. B}\ }\textbf {\bibinfo {volume} {86}},\ \bibinfo {pages} {134415} (\bibinfo {year} {2012})}\BibitemShut {NoStop}%
\bibitem [{\citenamefont {Jiang}\ \emph {et~al.}(2022)\citenamefont {Jiang}, \citenamefont {Xuan},\ and\ \citenamefont {Yu}}]{jiang_current-driven_2022}%
  \BibitemOpen
  \bibfield  {author} {\bibinfo {author} {\bibfnamefont {Y.}~\bibnamefont {Jiang}}, \bibinfo {author} {\bibfnamefont {C.}~\bibnamefont {Xuan}},\ and\ \bibinfo {author} {\bibfnamefont {H.}~\bibnamefont {Yu}},\ }\bibfield  {title} {\bibinfo {title} {Current-driven dynamics of skyrmions in the presence of pinning at finite temperatures},\ }\href {https://doi.org/10.1016/j.jmmm.2022.169786} {\bibfield  {journal} {\bibinfo  {journal} {J. Magn. Magn. Mater}\ }\textbf {\bibinfo {volume} {562}},\ \bibinfo {pages} {169786} (\bibinfo {year} {2022})}\BibitemShut {NoStop}%
\bibitem [{\citenamefont {Morgan}(2010)}]{morgan_surface_2010}%
  \BibitemOpen
  \bibfield  {author} {\bibinfo {author} {\bibfnamefont {D.~P.}\ \bibnamefont {Morgan}},\ }\href@noop {} {\emph {\bibinfo {title} {Surface {Acoustic} {Wave} {Filters}: {With} {Applications} to {Electronic} {Communications} and {Signal} {Processing}}}},\ \bibinfo {edition} {2nd}\ ed.,\ Studies in {Electrical} and {Electronic} {Engineering} {Ser}\ (\bibinfo  {publisher} {Elsevier Science \& Technology},\ \bibinfo {address} {San Diego},\ \bibinfo {year} {2010})\BibitemShut {NoStop}%
\bibitem [{\citenamefont {Weiler}\ \emph {et~al.}(2011)\citenamefont {Weiler}, \citenamefont {Dreher}, \citenamefont {Heeg}, \citenamefont {Huebl}, \citenamefont {Gross}, \citenamefont {Brandt},\ and\ \citenamefont {Goennenwein}}]{weiler_elastically_2011}%
  \BibitemOpen
  \bibfield  {author} {\bibinfo {author} {\bibfnamefont {M.}~\bibnamefont {Weiler}}, \bibinfo {author} {\bibfnamefont {L.}~\bibnamefont {Dreher}}, \bibinfo {author} {\bibfnamefont {C.}~\bibnamefont {Heeg}}, \bibinfo {author} {\bibfnamefont {H.}~\bibnamefont {Huebl}}, \bibinfo {author} {\bibfnamefont {R.}~\bibnamefont {Gross}}, \bibinfo {author} {\bibfnamefont {M.~S.}\ \bibnamefont {Brandt}},\ and\ \bibinfo {author} {\bibfnamefont {S.~T.~B.}\ \bibnamefont {Goennenwein}},\ }\bibfield  {title} {\bibinfo {title} {Elastically {Driven} {Ferromagnetic} {Resonance} in {Nickel} {Thin} {Films}},\ }\href {https://doi.org/10.1103/PhysRevLett.106.117601} {\bibfield  {journal} {\bibinfo  {journal} {Phys. Rev. Lett.}\ }\textbf {\bibinfo {volume} {106}},\ \bibinfo {pages} {117601} (\bibinfo {year} {2011})}\BibitemShut {NoStop}%
\bibitem [{\citenamefont {Weiler}\ \emph {et~al.}(2012)\citenamefont {Weiler}, \citenamefont {Huebl}, \citenamefont {Goerg}, \citenamefont {Czeschka}, \citenamefont {Gross},\ and\ \citenamefont {Goennenwein}}]{weiler_spin_2012}%
  \BibitemOpen
  \bibfield  {author} {\bibinfo {author} {\bibfnamefont {M.}~\bibnamefont {Weiler}}, \bibinfo {author} {\bibfnamefont {H.}~\bibnamefont {Huebl}}, \bibinfo {author} {\bibfnamefont {F.~S.}\ \bibnamefont {Goerg}}, \bibinfo {author} {\bibfnamefont {F.~D.}\ \bibnamefont {Czeschka}}, \bibinfo {author} {\bibfnamefont {R.}~\bibnamefont {Gross}},\ and\ \bibinfo {author} {\bibfnamefont {S.~T.~B.}\ \bibnamefont {Goennenwein}},\ }\bibfield  {title} {\bibinfo {title} {Spin {Pumping} with {Coherent} {Elastic} {Waves}},\ }\href {https://doi.org/10.1103/PhysRevLett.108.176601} {\bibfield  {journal} {\bibinfo  {journal} {Phys. Rev. Lett.}\ }\textbf {\bibinfo {volume} {108}},\ \bibinfo {pages} {176601} (\bibinfo {year} {2012})}\BibitemShut {NoStop}%
\bibitem [{\citenamefont {Gruber}\ \emph {et~al.}(2023)\citenamefont {Gruber}, \citenamefont {Brems}, \citenamefont {Rothörl}, \citenamefont {Sparmann}, \citenamefont {Schmitt}, \citenamefont {Kononenko}, \citenamefont {Kammerbauer}, \citenamefont {Syskaki}, \citenamefont {Farago}, \citenamefont {Virnau},\ and\ \citenamefont {Kläui}}]{gruber_300-times-increased_2023}%
  \BibitemOpen
  \bibfield  {author} {\bibinfo {author} {\bibfnamefont {R.}~\bibnamefont {Gruber}}, \bibinfo {author} {\bibfnamefont {M.~A.}\ \bibnamefont {Brems}}, \bibinfo {author} {\bibfnamefont {J.}~\bibnamefont {Rothörl}}, \bibinfo {author} {\bibfnamefont {T.}~\bibnamefont {Sparmann}}, \bibinfo {author} {\bibfnamefont {M.}~\bibnamefont {Schmitt}}, \bibinfo {author} {\bibfnamefont {I.}~\bibnamefont {Kononenko}}, \bibinfo {author} {\bibfnamefont {F.}~\bibnamefont {Kammerbauer}}, \bibinfo {author} {\bibfnamefont {M.-A.}\ \bibnamefont {Syskaki}}, \bibinfo {author} {\bibfnamefont {O.}~\bibnamefont {Farago}}, \bibinfo {author} {\bibfnamefont {P.}~\bibnamefont {Virnau}},\ and\ \bibinfo {author} {\bibfnamefont {M.}~\bibnamefont {Kläui}},\ }\bibfield  {title} {\bibinfo {title} {300-{Times}-{Increased} {Diffusive} {Skyrmion} {Dynamics} and {Effective} {Pinning} {Reduction} by {Periodic} {Field} {Excitation}},\ }\href {https://doi.org/10.1002/adma.202208922} {\bibfield  {journal} {\bibinfo  {journal} {Adv. Mater}\ }\textbf
  {\bibinfo {volume} {35}},\ \bibinfo {pages} {2208922} (\bibinfo {year} {2023})}\BibitemShut {NoStop}%
\bibitem [{\citenamefont {Vansteenkiste}\ \emph {et~al.}(2014)\citenamefont {Vansteenkiste}, \citenamefont {Leliaert}, \citenamefont {Dvornik}, \citenamefont {Helsen}, \citenamefont {Garcia-Sanchez},\ and\ \citenamefont {Van~Waeyenberge}}]{vansteenkiste_design_2014}%
  \BibitemOpen
  \bibfield  {author} {\bibinfo {author} {\bibfnamefont {A.}~\bibnamefont {Vansteenkiste}}, \bibinfo {author} {\bibfnamefont {J.}~\bibnamefont {Leliaert}}, \bibinfo {author} {\bibfnamefont {M.}~\bibnamefont {Dvornik}}, \bibinfo {author} {\bibfnamefont {M.}~\bibnamefont {Helsen}}, \bibinfo {author} {\bibfnamefont {F.}~\bibnamefont {Garcia-Sanchez}},\ and\ \bibinfo {author} {\bibfnamefont {B.}~\bibnamefont {Van~Waeyenberge}},\ }\bibfield  {title} {\bibinfo {title} {The design and verification of {MuMax3}},\ }\href {https://doi.org/10.1063/1.4899186} {\bibfield  {journal} {\bibinfo  {journal} {AIP Adv.}\ }\textbf {\bibinfo {volume} {4}},\ \bibinfo {pages} {107133} (\bibinfo {year} {2014})}\BibitemShut {NoStop}%
\bibitem [{noa()}]{noauthor_see_nodate}%
  \BibitemOpen
  \href {www.aithericon.com} {\bibinfo {title} {See www.aithericon.com for {Aithericon}}}\BibitemShut {NoStop}%
\bibitem [{\citenamefont {Kim}\ \emph {et~al.}(2018)\citenamefont {Kim}, \citenamefont {Moon}, \citenamefont {Kerber}, \citenamefont {Nothhelfer},\ and\ \citenamefont {Everschor-Sitte}}]{kim_asymmetric_2018}%
  \BibitemOpen
  \bibfield  {author} {\bibinfo {author} {\bibfnamefont {K.-W.}\ \bibnamefont {Kim}}, \bibinfo {author} {\bibfnamefont {K.-W.}\ \bibnamefont {Moon}}, \bibinfo {author} {\bibfnamefont {N.}~\bibnamefont {Kerber}}, \bibinfo {author} {\bibfnamefont {J.}~\bibnamefont {Nothhelfer}},\ and\ \bibinfo {author} {\bibfnamefont {K.}~\bibnamefont {Everschor-Sitte}},\ }\bibfield  {title} {\bibinfo {title} {Asymmetric skyrmion {Hall} effect in systems with a hybrid {Dzyaloshinskii}-{Moriya} interaction},\ }\href {https://doi.org/10.1103/PhysRevB.97.224427} {\bibfield  {journal} {\bibinfo  {journal} {Phys. Rev. B}\ }\textbf {\bibinfo {volume} {97}},\ \bibinfo {pages} {224427} (\bibinfo {year} {2018})}\BibitemShut {NoStop}%
\bibitem [{\citenamefont {Woo}\ \emph {et~al.}(2018)\citenamefont {Woo}, \citenamefont {Song}, \citenamefont {Zhang}, \citenamefont {Zhou}, \citenamefont {Ezawa}, \citenamefont {Liu}, \citenamefont {Finizio}, \citenamefont {Raabe}, \citenamefont {Lee}, \citenamefont {Kim}, \citenamefont {Park}, \citenamefont {Kim}, \citenamefont {Kim}, \citenamefont {Lee}, \citenamefont {Lee}, \citenamefont {Choi}, \citenamefont {Min}, \citenamefont {Koo},\ and\ \citenamefont {Chang}}]{woo_current-driven_2018}%
  \BibitemOpen
  \bibfield  {author} {\bibinfo {author} {\bibfnamefont {S.}~\bibnamefont {Woo}}, \bibinfo {author} {\bibfnamefont {K.~M.}\ \bibnamefont {Song}}, \bibinfo {author} {\bibfnamefont {X.}~\bibnamefont {Zhang}}, \bibinfo {author} {\bibfnamefont {Y.}~\bibnamefont {Zhou}}, \bibinfo {author} {\bibfnamefont {M.}~\bibnamefont {Ezawa}}, \bibinfo {author} {\bibfnamefont {X.}~\bibnamefont {Liu}}, \bibinfo {author} {\bibfnamefont {S.}~\bibnamefont {Finizio}}, \bibinfo {author} {\bibfnamefont {J.}~\bibnamefont {Raabe}}, \bibinfo {author} {\bibfnamefont {N.~J.}\ \bibnamefont {Lee}}, \bibinfo {author} {\bibfnamefont {S.-I.}\ \bibnamefont {Kim}}, \bibinfo {author} {\bibfnamefont {S.-Y.}\ \bibnamefont {Park}}, \bibinfo {author} {\bibfnamefont {Y.}~\bibnamefont {Kim}}, \bibinfo {author} {\bibfnamefont {J.-Y.}\ \bibnamefont {Kim}}, \bibinfo {author} {\bibfnamefont {D.}~\bibnamefont {Lee}}, \bibinfo {author} {\bibfnamefont {O.}~\bibnamefont {Lee}}, \bibinfo {author} {\bibfnamefont {J.~W.}\ \bibnamefont {Choi}}, \bibinfo {author}
  {\bibfnamefont {B.-C.}\ \bibnamefont {Min}}, \bibinfo {author} {\bibfnamefont {H.~C.}\ \bibnamefont {Koo}},\ and\ \bibinfo {author} {\bibfnamefont {J.}~\bibnamefont {Chang}},\ }\bibfield  {title} {\bibinfo {title} {Current-driven dynamics and inhibition of the skyrmion {Hall} effect of ferrimagnetic skyrmions in {GdFeCo} films},\ }\href {https://doi.org/10.1038/s41467-018-03378-7} {\bibfield  {journal} {\bibinfo  {journal} {Nat. Commun.}\ }\textbf {\bibinfo {volume} {9}},\ \bibinfo {pages} {959} (\bibinfo {year} {2018})}\BibitemShut {NoStop}%
\bibitem [{\citenamefont {Brearton}\ \emph {et~al.}(2021)\citenamefont {Brearton}, \citenamefont {Turnbull}, \citenamefont {Verezhak}, \citenamefont {Balakrishnan}, \citenamefont {Hatton}, \citenamefont {van~der Laan},\ and\ \citenamefont {Hesjedal}}]{brearton_deriving_2021}%
  \BibitemOpen
  \bibfield  {author} {\bibinfo {author} {\bibfnamefont {R.}~\bibnamefont {Brearton}}, \bibinfo {author} {\bibfnamefont {L.~A.}\ \bibnamefont {Turnbull}}, \bibinfo {author} {\bibfnamefont {J.~a.~T.}\ \bibnamefont {Verezhak}}, \bibinfo {author} {\bibfnamefont {G.}~\bibnamefont {Balakrishnan}}, \bibinfo {author} {\bibfnamefont {P.~D.}\ \bibnamefont {Hatton}}, \bibinfo {author} {\bibfnamefont {G.}~\bibnamefont {van~der Laan}},\ and\ \bibinfo {author} {\bibfnamefont {T.}~\bibnamefont {Hesjedal}},\ }\bibfield  {title} {\bibinfo {title} {Deriving the skyrmion {Hall} angle from skyrmion lattice dynamics},\ }\href {https://doi.org/10.1038/s41467-021-22857-y} {\bibfield  {journal} {\bibinfo  {journal} {Nat. Commun.}\ }\textbf {\bibinfo {volume} {12}},\ \bibinfo {pages} {2723} (\bibinfo {year} {2021})}\BibitemShut {NoStop}%
\bibitem [{\citenamefont {Jiang}\ \emph {et~al.}(2017)\citenamefont {Jiang}, \citenamefont {Zhang}, \citenamefont {Yu}, \citenamefont {Zhang}, \citenamefont {Wang}, \citenamefont {Benjamin~Jungfleisch}, \citenamefont {Pearson}, \citenamefont {Cheng}, \citenamefont {Heinonen}, \citenamefont {Wang}, \citenamefont {Zhou}, \citenamefont {Hoffmann},\ and\ \citenamefont {te~Velthuis}}]{jiang_direct_2017}%
  \BibitemOpen
  \bibfield  {author} {\bibinfo {author} {\bibfnamefont {W.}~\bibnamefont {Jiang}}, \bibinfo {author} {\bibfnamefont {X.}~\bibnamefont {Zhang}}, \bibinfo {author} {\bibfnamefont {G.}~\bibnamefont {Yu}}, \bibinfo {author} {\bibfnamefont {W.}~\bibnamefont {Zhang}}, \bibinfo {author} {\bibfnamefont {X.}~\bibnamefont {Wang}}, \bibinfo {author} {\bibfnamefont {M.}~\bibnamefont {Benjamin~Jungfleisch}}, \bibinfo {author} {\bibfnamefont {J.~E.}\ \bibnamefont {Pearson}}, \bibinfo {author} {\bibfnamefont {X.}~\bibnamefont {Cheng}}, \bibinfo {author} {\bibfnamefont {O.}~\bibnamefont {Heinonen}}, \bibinfo {author} {\bibfnamefont {K.~L.}\ \bibnamefont {Wang}}, \bibinfo {author} {\bibfnamefont {Y.}~\bibnamefont {Zhou}}, \bibinfo {author} {\bibfnamefont {A.}~\bibnamefont {Hoffmann}},\ and\ \bibinfo {author} {\bibfnamefont {S.~G.~E.}\ \bibnamefont {te~Velthuis}},\ }\bibfield  {title} {\bibinfo {title} {Direct observation of the skyrmion {Hall} effect},\ }\href {https://doi.org/10.1038/nphys3883} {\bibfield  {journal}
  {\bibinfo  {journal} {Nat. Phys.}\ }\textbf {\bibinfo {volume} {13}},\ \bibinfo {pages} {162} (\bibinfo {year} {2017})}\BibitemShut {NoStop}%
\bibitem [{\citenamefont {Yang}\ \emph {et~al.}(2024{\natexlab{b}})\citenamefont {Yang}, \citenamefont {Zhao}, \citenamefont {Zhang}, \citenamefont {Xing}, \citenamefont {Du}, \citenamefont {Li}, \citenamefont {Mochizuki}, \citenamefont {Xu}, \citenamefont {Åkerman},\ and\ \citenamefont {Zhou}}]{yang_fundamentals_2024}%
  \BibitemOpen
  \bibfield  {author} {\bibinfo {author} {\bibfnamefont {S.}~\bibnamefont {Yang}}, \bibinfo {author} {\bibfnamefont {Y.}~\bibnamefont {Zhao}}, \bibinfo {author} {\bibfnamefont {X.}~\bibnamefont {Zhang}}, \bibinfo {author} {\bibfnamefont {X.}~\bibnamefont {Xing}}, \bibinfo {author} {\bibfnamefont {H.}~\bibnamefont {Du}}, \bibinfo {author} {\bibfnamefont {X.}~\bibnamefont {Li}}, \bibinfo {author} {\bibfnamefont {M.}~\bibnamefont {Mochizuki}}, \bibinfo {author} {\bibfnamefont {X.}~\bibnamefont {Xu}}, \bibinfo {author} {\bibfnamefont {J.}~\bibnamefont {Åkerman}},\ and\ \bibinfo {author} {\bibfnamefont {Y.}~\bibnamefont {Zhou}},\ }\bibfield  {title} {\bibinfo {title} {Fundamentals and applications of the skyrmion {Hall} effect},\ }\href {https://doi.org/10.1063/5.0218280} {\bibfield  {journal} {\bibinfo  {journal} {Appl. Phys. Rev.}\ }\textbf {\bibinfo {volume} {11}},\ \bibinfo {pages} {041335} (\bibinfo {year} {2024}{\natexlab{b}})}\BibitemShut {NoStop}%
\bibitem [{\citenamefont {Chen}(2017)}]{chen_skyrmion_2017}%
  \BibitemOpen
  \bibfield  {author} {\bibinfo {author} {\bibfnamefont {G.}~\bibnamefont {Chen}},\ }\bibfield  {title} {\bibinfo {title} {Skyrmion {Hall} effect},\ }\href {https://doi.org/10.1038/nphys4030} {\bibfield  {journal} {\bibinfo  {journal} {Nat. Phys.}\ }\textbf {\bibinfo {volume} {13}},\ \bibinfo {pages} {112} (\bibinfo {year} {2017})}\BibitemShut {NoStop}%
\bibitem [{\citenamefont {Litzius}\ \emph {et~al.}(2017)\citenamefont {Litzius}, \citenamefont {Lemesh}, \citenamefont {Krüger}, \citenamefont {Bassirian}, \citenamefont {Caretta}, \citenamefont {Richter}, \citenamefont {Büttner}, \citenamefont {Sato}, \citenamefont {Tretiakov}, \citenamefont {Förster}, \citenamefont {Reeve}, \citenamefont {Weigand}, \citenamefont {Bykova}, \citenamefont {Stoll}, \citenamefont {Schütz}, \citenamefont {Beach},\ and\ \citenamefont {Kläui}}]{litzius_skyrmion_2017}%
  \BibitemOpen
  \bibfield  {author} {\bibinfo {author} {\bibfnamefont {K.}~\bibnamefont {Litzius}}, \bibinfo {author} {\bibfnamefont {I.}~\bibnamefont {Lemesh}}, \bibinfo {author} {\bibfnamefont {B.}~\bibnamefont {Krüger}}, \bibinfo {author} {\bibfnamefont {P.}~\bibnamefont {Bassirian}}, \bibinfo {author} {\bibfnamefont {L.}~\bibnamefont {Caretta}}, \bibinfo {author} {\bibfnamefont {K.}~\bibnamefont {Richter}}, \bibinfo {author} {\bibfnamefont {F.}~\bibnamefont {Büttner}}, \bibinfo {author} {\bibfnamefont {K.}~\bibnamefont {Sato}}, \bibinfo {author} {\bibfnamefont {O.~A.}\ \bibnamefont {Tretiakov}}, \bibinfo {author} {\bibfnamefont {J.}~\bibnamefont {Förster}}, \bibinfo {author} {\bibfnamefont {R.~M.}\ \bibnamefont {Reeve}}, \bibinfo {author} {\bibfnamefont {M.}~\bibnamefont {Weigand}}, \bibinfo {author} {\bibfnamefont {I.}~\bibnamefont {Bykova}}, \bibinfo {author} {\bibfnamefont {H.}~\bibnamefont {Stoll}}, \bibinfo {author} {\bibfnamefont {G.}~\bibnamefont {Schütz}}, \bibinfo {author} {\bibfnamefont {G.~S.~D.}\
  \bibnamefont {Beach}},\ and\ \bibinfo {author} {\bibfnamefont {M.}~\bibnamefont {Kläui}},\ }\bibfield  {title} {\bibinfo {title} {Skyrmion {Hall} effect revealed by direct time-resolved {X}-ray microscopy},\ }\href {https://doi.org/10.1038/nphys4000} {\bibfield  {journal} {\bibinfo  {journal} {Nat. Phys.}\ }\textbf {\bibinfo {volume} {13}},\ \bibinfo {pages} {170} (\bibinfo {year} {2017})}\BibitemShut {NoStop}%
\bibitem [{sup()}]{supplement}%
  \BibitemOpen
  \href@noop {} {\bibinfo {title} {See supplemental material at [url] for detailed information on the skyrmion motion over time when applying a sinusoidal and sawtooth-shaped saw to the skyrmion in a chain pinning and more realistic random grain pinning landscape.}}\BibitemShut {Stop}%
\bibitem [{\citenamefont {Ashby}(2011)}]{ashby_chapter_2011}%
  \BibitemOpen
  \bibfield  {author} {\bibinfo {author} {\bibfnamefont {M.~F.}\ \bibnamefont {Ashby}},\ }\bibfield  {title} {\bibinfo {title} {Chapter 4 - {Material} {Property} {Charts}},\ }in\ \href {https://doi.org/10.1016/B978-1-85617-663-7.00004-7} {\emph {\bibinfo {booktitle} {Materials {Selection} in {Mechanical} {Design} ({Fourth} {Edition})}}},\ \bibinfo {editor} {edited by\ \bibinfo {editor} {\bibfnamefont {M.~F.}\ \bibnamefont {Ashby}}}\ (\bibinfo  {publisher} {Butterworth-Heinemann},\ \bibinfo {address} {Oxford},\ \bibinfo {year} {2011})\ pp.\ \bibinfo {pages} {57--96}\BibitemShut {NoStop}%
\bibitem [{\citenamefont {Kavalerov}\ \emph {et~al.}(2000)\citenamefont {Kavalerov}, \citenamefont {Fujii},\ and\ \citenamefont {Inoue}}]{kavalerov_observation_2000}%
  \BibitemOpen
  \bibfield  {author} {\bibinfo {author} {\bibfnamefont {V.}~\bibnamefont {Kavalerov}}, \bibinfo {author} {\bibfnamefont {T.}~\bibnamefont {Fujii}},\ and\ \bibinfo {author} {\bibfnamefont {M.}~\bibnamefont {Inoue}},\ }\bibfield  {title} {\bibinfo {title} {Observation of highly nonlinear surface-acoustic waves on single crystal lithium–niobate plates by means of an optical sampling probe},\ }\href {https://doi.org/10.1063/1.371960} {\bibfield  {journal} {\bibinfo  {journal} {J. Appl. Phys.}\ }\textbf {\bibinfo {volume} {87}},\ \bibinfo {pages} {907} (\bibinfo {year} {2000})}\BibitemShut {NoStop}%
\bibitem [{\citenamefont {Schwenke}\ \emph {et~al.}(2025)\citenamefont {Schwenke}, \citenamefont {Spindler}, \citenamefont {Vasyuchka}, \citenamefont {Hamadeh}, \citenamefont {Pirro},\ and\ \citenamefont {Weiler}}]{schwenke_ratchet_2025}%
  \BibitemOpen
  \bibfield  {author} {\bibinfo {author} {\bibfnamefont {P.}~\bibnamefont {Schwenke}}, \bibinfo {author} {\bibfnamefont {E.}~\bibnamefont {Spindler}}, \bibinfo {author} {\bibfnamefont {V.}~\bibnamefont {Vasyuchka}}, \bibinfo {author} {\bibfnamefont {A.}~\bibnamefont {Hamadeh}}, \bibinfo {author} {\bibfnamefont {P.}~\bibnamefont {Pirro}},\ and\ \bibinfo {author} {\bibfnamefont {M.}~\bibnamefont {Weiler}},\ }\href {https://doi.org/10.5281/zenodo.15019342} {\bibinfo {title} {Zenodo}} (\bibinfo {year} {2025})\BibitemShut {NoStop}%
\end{thebibliography}

%

\end{document}